%% file: ccfm.tex
\documentclass[10pt,draftcls,onecolumn]{IEEEtran}
\usepackage{amsmath}
\usepackage{amssymb}
\usepackage{upgreek}
\usepackage{tipa}
\usepackage{enumerate}
\usepackage{subfig}
\usepackage{float}
\usepackage{multicol}
\usepackage[usenames,dvipsnames]{pstricks}
\usepackage{epsfig}
\usepackage{pst-grad} 
\usepackage{pst-plot} 
\usepackage[noadjust]{cite}
\usepackage{caption}
\usepackage{graphicx}
\usepackage{epstopdf}
\usepackage{psfrag}
\usepackage{amsfonts}
\usepackage{color}
\usepackage{amsmath,mathtools}
\usepackage{psfrag}
\usepackage{comment}
\usepackage{url}
\usepackage{balance}

\begin{document}
\title{Stability, convergence and Hopf bifurcation analyses of the classical car-following model}

\author{\IEEEauthorblockN{Gopal Krishna Kamath$^{\ast}$, Krishna Jagannathan and Gaurav Raina} \\
\IEEEauthorblockA{Department of Electrical Engineering, Indian Institute of Technology Madras, Chennai 600 036, India\\
Email: $\lbrace \text{ee12d033, krishnaj, gaurav} \rbrace$@ee.iitm.ac.in}
\thanks{$^{\ast}$ Corresponding author}
\thanks{This is an extension of our preliminary work that appeared in Proceedings of the $53^{rd}$ Annual Allerton Conference on Communication, Control and Computing, pp. 538-545, 2015. DOI: 10.1109/ALLERTON.2015.7447051}
}
\maketitle

\begin{abstract}
Reaction delays play an important role in determining the qualitative dynamical properties of a platoon of vehicles traversing a straight road. In this paper, we investigate the impact of delayed feedback on the dynamics of the Classical Car-Following Model (CCFM). Specifically, we analyze the CCFM in no delay, small delay and arbitrary delay regimes. First, we derive a sufficient condition for local stability of the CCFM in no-delay and small-delay regimes using. Next, we derive the necessary and sufficient condition for local stability of the CCFM for an arbitrary delay. We then demonstrate that the transition of traffic flow from the locally stable to the unstable regime occurs via a Hopf bifurcation, thus resulting in limit cycles in system dynamics. Physically, these limit cycles manifest as back-propagating congestion waves on highways.

In the context of human-driven vehicles, our work provides phenomenological insight into the impact of reaction delays on the emergence and evolution of traffic congestion. In the context of self-driven vehicles, our work has the potential to provide design guidelines for control algorithms running in self-driven cars to avoid undesirable phenomena. Specifically, designing control algorithms that avoid jerky vehicular movements is essential. Hence, we derive the necessary and sufficient condition for non-oscillatory convergence of the CCFM. Next, we characterize the rate of convergence of the CCFM, and bring forth the interplay between local stability, non-oscillatory convergence and the rate of convergence of the CCFM.

Further, to better understand the oscillations in the system dynamics, we characterize the type of the Hopf bifurcation and the asymptotic orbital stability of the limit cycles using Poincar\'{e} normal forms and the center manifold theory. The analysis is complemented with stability charts, bifurcation diagrams and MATLAB simulations.

\end{abstract}


\IEEEpeerreviewmaketitle

\section{Introduction}
\label{sec:intro}
\input{sec1_intro.tex}

\section{Models}
\label{sec:models}
\input{sec2_models.tex}

\section{The no-delay regime}
\label{sec:ideal}
\input{sec3_ideal.tex}

\section{The small-delay regime}
\label{sec:small}
\input{sec4_small.tex}

\section{The Hopf bifurcation}
\label{sec:hopf}
\input{sec5_hopf.tex}

\section{Non-oscillatory convergence}
\label{sec:noc}
\input{sec6_noc.tex}

\section{Rate of convergence}
\label{sec:roc}
\input{sec7_roc.tex}

\section{Hopf bifurcation analysis}
\label{section:RCCFMhopf}
\input{sec8_hba.tex}

\section{Simulations}
\label{sec:sims}
\input{sec9_sims.tex}

\section{Concluding Remarks}
\label{sec:conc}
\input{sec10_conc.tex}

\section*{Acknowledgements}
This work is undertaken as a part of an Information Technology Research Academy (ITRA), Media Lab Asia, project titled ``De-congesting India's transportation networks.'' The authors are also thankful to Debayani Ghosh and Sreelakshmi Manjunath for many helpful discussions.

\end{document}

%% file: sec1_intro.tex
Intelligent transportation systems constitute a substantial theme of discussion on futuristic smart cities. A prospective solution to increase resource utilization is to use self-driven vehicles, which may also mitigate traffic congestion~\cite[Section 5.2]{RR},~\cite{SG}. To that end, it is imperative to design stable control algorithms for these vehicles. Since a good design process requires an in-depth understanding of vehicular dynamics, a class of dynamical models -- known as car-following models -- have been developed and studied~\cite{DCG, DC, DH, GKK, GO, MBD, XZ}.

An important consideration in the study of car-following models is the delay in the dynamical variables. Delays arise due to various factors such as sensing, mechanical motions, communication  and signal processing. These delays are known to have a variety of effects on the properties of a dynamical system~\cite{HL}. Specifically, delays can readily lead to oscillations and instability~\cite{RS, XZ}.

In this paper, we investigate the impact of delayed feedback on the qualitative dynamical properties of a platoon of vehicles driving on a straight road. Specifically, we focus on analyzing the effect of delayed feedback on the Classical Car-Following Model (CCFM). In the specific context of human-driven vehicles, the dominant sources of delay are the physiological delay and the mechanical delay~\cite{RS}. In contrast, self-driven vehicles tend to have smaller reaction delays than their human-driven counterparts, and typically occur due to the delays in sensing, computation and actuation~\cite{AK}. Hence, we analyze local stability of the CCFM in three regimes -- no delay, small delay and arbitrary delay.

In addition to stability, non-oscillatory convergence and rate of convergence constitute two properties of practical interest, which we also explore for the case of the CCFM. Such conditions could aid in ensuring smooth traffic flow by avoiding jerky vehicular motion, thereby improving ride quality. The theoretical analyses could offer suggestions for design guidelines.


In the context of human-driven vehicles, our investigation into the impact of reaction delay enhances phenomenological insights into the emergence and evolution of traffic congestion. For example, a peculiar phenomenon known as a `phantom jam' -- the emergence of a back-propagating congestion wave in motorway traffic, seemingly out of nowhere -- has been observed in the real world~\cite{DC,DH}. Previous studies~\cite{DC,DH} have shown that a change in driver's sensitivity (for instance, a sudden deceleration) can lead to such oscillatory behaviour. In this paper, we show that similar oscillations could also result from an increase in the driver's reaction delay. More generally, our study leads to an important observation that the transition of traffic flow from stability to instability could take place due to a variation in many combinations of model parameters. In order to capture this complex dependence on various parameters, we introduce an exogenous, non-dimensional parameter in our dynamical model, set to unity on the stability boundary. We then analyze the system behavior as this exogenous parameter pushes the system across the stability boundary, and show that limit cycles emerge due to a Hopf bifurcation.

The impact of the reaction delay is perhaps even more important in the context of self-driven vehicles. Self-driven vehicles are envisioned to have reduced reaction delays as compared to a human driver. As a result, self-driven vehicles facilitate smaller equilibrium separation between consecutive vehicles~\cite[Section 5.2]{RR}. This, in turn, improves resource utilization without compromising safety~\cite{SG}.  In contrast to the case of human-driven vehicles, the parameters in the control algorithm -- known as upper longitudinal control algorithm~\cite[Section 5.2]{RR} -- for self-driven vehicles need to be tuned appropriately. To that end, our analyses and findings highlight the quantitative impact of delayed feedback on the design of control algorithms for self-driven vehicles. In particular, the combination of stability and convergence analyses may help in the design of various aspects of longitudinal control algorithms~\cite[Section 5.2]{RR}. We complement our theoretical analyses using stability charts, bifurcation diagrams and MATLAB simulations.

\subsection{Related work on car-following models}   

The work by Chandler \emph{et al.}~\cite{REC} as well as the one by Herman \emph{et al.}~\cite{RH} constitute two of the earliest known investigations on stability of car-following models. The CCFM was proposed in~\cite{DCG}, although the main objective therein was to understand the resulting macroscopic behavior. Several related models, and their modifications, have been investigated in~\cite{EAU},~\cite{REC} and~\cite{RH}. For a recent exposition of linear stability analysis as applied to car-following models, see~\cite{REW}. The aforementioned investigations mainly use transform techniques to derive conditions for stability.

In contrast,~\cite{XZ} and some of the references therein consider the issue of stability from a dynamical systems perspective. Specifically,~\cite{XZ} studies some stability properties of the CCFM. However, the aforementioned works do not consider the delay in the the self-velocity term. To make the model more realistic, we accounted for this delay in our previous work~\cite{GKK}. Therein, we studied a particular case of the CCFM called the Reduced Classical Car-Following Model (RCCFM), and showed that it loses local stability via a Hopf bifurcation. This paper extends the results presented in~\cite{GKK} to the CCFM, and also derives conditions that may ensure good ride quality, in addition to characterizing the time taken by a platoon to reach its equilibrium. To the best of our knowledge, ours is the first work to characterize such a metric. Further, we show that oscillations in state variables are a manifestation of \emph{limit cycles}, and not centers as asserted in~\cite{XZ}. For a recent review on stability analyses as applied to car-following models, see~\cite{JZJ}. For an exposition on the use of time-delayed equations for traffic-flow modeling, see~\cite{RS1}.

Note that several dynamical models have also been studied in the Physics literature beginning with the Optimal Velocity Model (OVM)~\cite{MBD}. In fact, it is known that some of these models lose local stability via a Hopf bifurcation as well~\cite{GO2, GO, IG}. The OVM has also been studied as a Fillipov system by interpreting negative inter-vehicular distance as an overtaking maneuver~\cite{LB}. Further, macroscopic traffic jams resulting due to the OVM have been studied using the Korteweg-de Vries equation; see~\cite{LH} and references therein for details. However, this body of literature assumes the vehicles to be traveling on a single-lane circular loop, thus mathematically yielding periodic boundary conditions. In contrast, the CCFM and related models differ at a fundamental level by assuming the vehicular motion on a single-lane straight road. Thus, we do not attempt to compare our results with those derived for the OVM and related models.

From a vehicular dynamics perspective, most upper longitudinal controllers in the literature assume the lower controller's dynamics to be well-modeled by a first-order control system, in order to capture the delay lag~\cite[Section 5.3]{RR}. The upper longitudinal controllers are then designed to maintain either constant velocity, spacing or time gap; for details, see~\cite{RR1} and the references therein. Specifically, Rajamani \emph{et al.}~\cite{RR1} prove that synchronization with the lead vehicle is possible by using information only from the vehicle directly ahead. This reduces implementation complexity, and does not mandate vehicles to be installed with communication devices.

However, in the context of autonomous vehicles, communication systems are required to exchange various system states required for the control algorithm. This information is used either for distributed control~\cite{RR1} or coordinated control~\cite{ZQ}. Formation and platoon stabilities have also been studied considering information flow among the vehicles~\cite{THS, RUC}. For an extensive review, see~\cite{KCD}.

In contrast to stabilizing platoons of autonomous vehicles (our scenario), it has been shown that well-placed, communicating autonomous vehicles may be used to stabilize platoons of human-driven vehicles as well~\cite{GO1}. More generally, the platooning problem has been studied as a consensus problem with delays~\cite{MAS}. Such an approach aids the design of coupling protocols between interacting agents (in this context, vehicles). In contrast, we provide design guidelines to appropriately choose protocol parameters, given a coupling protocol (the CCFM).


\subsection{Our contributions}
Our contributions can be summarized as follows.
\begin{itemize}
\item[(1)] We make the CCFM more realistic by accounting for the delay in the self-velocity term.
\item[(2)] We show that, in the absence of reaction delays, the CCFM is locally stable for all parameter values of practical interest. When the delays are rather small, we derive a sufficient condition for local stability of the CCFM using a linearization of the time variable.
\item[(3)] We derive the necessary and sufficient condition for the local stability of the CCFM for an arbitrary delay. We then show that, upon violation of this condition, the CCFM loses local stability via a Hopf bifurcation. Indeed, this helps us understand that the oscillations emerge as a consequence of limit cycles, and centers as asserted in the literature.
\item[(4)] In the case of human-driven vehicles, our work enhances phenomenological insights into the emergence and evolution of traffic congestion. For example, the notion of Hopf bifurcation provides a mathematical framework to offer a possible explanation for the observed `phantom jams.'
\item[(5)] We derive the necessary and sufficient condition for non-oscillatory convergence of the CCFM. This is useful in the context of a transportation network since oscillations lead to jerky vehicular movements, thereby degrading ride quality and possibly causing collisions.
\item[(6)] We characterize the rate of convergence of the CCFM, thereby gaining insight into the time required for the platoon to attain the desired equilibrium, when perturbed. Such perturbations occur, for instance, when a vehicle departs from a platoon.
\item[(7)] We highlight the three-way trade-off between local stability, non-oscillatory convergence and the rate of convergence. Considering this trade-off, we suggest some guidelines to appropriately choose parameters for the upper longitudinal control algorithm in self-driven vehicles.
\item[(8)] We characterize the type of Hopf bifurcation and the asymptotic orbital stability of the emergent limit cycles using  Poincar\'{e} normal forms and the center manifold theory.
\item[(9)] We corroborate the analytical results with the aid of stability charts, numerical computations and simulations conducted using MATLAB.
\end{itemize}



The remainder of this paper is organised as follows. In Section~\ref{sec:models}, we introduce the CCFM. In Sections~\ref{sec:ideal},~\ref{sec:small} and~\ref{sec:hopf}, we characterize the stable region for the CCFM in no-delay, small-delay and arbitrary-delay regimes respectively. We understand the stable region by characterizing the region of non-oscillatory convergence of the CCFM in Section~\ref{sec:noc}, and the rate of convergence of the CCFM in Section~\ref{sec:roc}. In Section~\ref{section:RCCFMhopf}, we present the local Hopf bifurcation analysis for the CCFM. In Section~\ref{sec:sims}, we present the simulation results before concluding in Section~\ref{sec:conc}.

%% file: sec2_models.tex
We begin this section with an overview of the setting of our work. We then briefly explain the CCFM.

\subsection{The setting}
We study a platoon of $N+1$ `ideal' ($i.e.,$ zero length) vehicles traversing an infinitely long, single-lane road without overtaking. The lead vehicle is indexed $0,$ its follower $1,$ and so forth. Each vehicle updates its acceleration based on a combination of its position, velocity and acceleration and those of the vehicle directly ahead. Let $x_i(t),$ $\dot{x}_i(t)$ and $\ddot{x}_i(t)$ denote the position, velocity and acceleration of the $i^{th}$ vehicle respectively, at time $t.$ The acceleration and velocity profiles of the lead vehicle are assumed to be known. In particular, we restrict ourselves to leader profiles that converge, in finite time, to $\ddot{x}_0 = 0$ and $0 < \text{ } \dot{x}_0 < \infty$; that is, there exists a finite $T_0$ such that $\ddot{x}_0(t) = 0,$ $\dot{x}_0(t) = \dot{x}_0 \, > \, 0,$ $\forall \, t \geq T_0$. We use the terms ``driver" and ``vehicle" interchangeably throughout. Further, we use SI units throughout.


\subsection{The Classical Car-Following Model (CCFM)}
A key feature of the CCFM is that the acceleration of each vehicle is dependent on three quantities: $(i)$ its own velocity, $(ii)$ velocity relative to the vehicle directly ahead, and $(iii)$ distance to the vehicle directly ahead. The exact dependence has been modeled in the literature as~\cite{DCG}
\begin{align}
\label{eq:CCFMO}
\ddot{x}_i(t) = \alpha_i \frac{ \left( \dot{x}_i(t) \right)^m \left( \dot{x}_{i-1}(t - \tau) - \dot{x}_{i}(t - \tau) \right)}{\left( x_{i-1}(t - \tau) - x_{i}(t - \tau)\right)^l},
\end{align}
for $i \in \lbrace 1, 2, \cdots, N \rbrace$. Here, $\alpha_i \, > \, 0$ represents the $i^{th}$ driver's sensitivity coefficient, for each $i \in \lbrace 1, 2, \cdots, N \rbrace$. Also, $m \in [-2,2]$ and $l \in \mathbb{R}_{+}$ are model parameters that contribute to the non-linearity. Note that the reaction delay is neglected in the self-velocity term $\left( \dot{x}_i(t) \right)^m.$ While self velocity might be available almost immediately, it takes some non-negligible time to execute the required control action. Also, from an analytical viewpoint, ignoring delays (in general) may generate inaccurate results. Thus, we account for the delay in the self-velocity term. Further, to make the model more realistic, we assume heterogeneity in reaction delays of different vehicles.


It is apparent from~\eqref{eq:CCFMO} that the state variable $x_i(t)$ becomes unbounded as $t \rightarrow \infty$ for each $i$. Therefore, similar to~\cite{XZ}, we transform the model in~\eqref{eq:CCFMO} using $y_i(t)$ + $b_i$ = $x_{i-1}(t) - x_i(t)$ and $v_i(t)$ = $\dot{y}_i(t)$ = $\dot{x}_{i-1}(t) - \dot{x}_i(t)$ for $i \in \lbrace 1, 2, \cdots, N \rbrace$. Here, $b_i$ denotes the desired equilibrium separation for the $i^{th}$ pair, $y_i(t) + b_i$ represents the separation between vehicles $i-1$ and $i$ at time $t,$ and $v_i(t)$ corresponds to the relative velocity of the $i^{th}$ vehicle with respect to the $(i-1)^{th}$ vehicle at time $t$. The transformed model is thus obtained as
\begin{align}
\nonumber
\dot{v}_i(t) = & \, \beta_{i-1}(t - \tau_{i-1}) v_{i-1}(t - \tau_{i-1}) - \beta_i( t - \tau_i)  v_{i}(t - \tau_i), \\ \label{eq:CCFMT12}
\dot{y}_i(t) = & \, v_i(t),
\end{align}
for $i \in \lbrace 1, 2, \cdots, N \rbrace$. Here,
\begin{align*}
\beta_i(t) = \, \alpha_{i} \frac{ \left( \dot{x}_0(t) - v_0(t) - \cdots - v_{i}(t) \right)^m}{\left( y_{i}(t) + b_i \right)^l}.
\end{align*}
Note that $y_0$, $v_0$, $\alpha_0$ and $\tau_0$ are dummy variables introduced for notational brevity, all of which are set to zero. We emphasize that $y_0$ and $v_0$ are \emph{not} state variables.

Note that $y_i(t) + b_i,$ and \emph{not} $y_i(t),$ represents the headway at time $t.$ In fact, $y_i(t)$ represents the variation of the headway about its equilibrium $b_i.$ Thus, $y_i(t)$ may become negative. However, the model breaks down when $y_i(t) + b_i$ becomes zero for $l > 0$~\cite{XZ}. Also, the CCFM possesses an inherent ``repulsion'' property, which may be illustrated as follows. Suppose that the vehicle indexed $i$ approaches the vehicle indexed $i-1$ at a relatively higher velocity. When the distance becomes very small (mathematically, $<1$ meter), the $i^{th}$ vehicle decelerates rather rapidly. This can be inferred from~\eqref{eq:CCFMO}. This helps avoid collision (hence the term ``repulsion''), thus ensuring $y_i(t) + b_i > 0.$

Since equations of the form~\eqref{eq:CCFMT12} are hard to analyze, we obtain sufficient conditions for their stability by analyzing them in the neighborhood of their equilibria. To that end, note that $v_i^* = 0,$ $y_i^* = 0$ $i = 1, 2, \cdots, N$ is an equilibrium for system~\eqref{eq:CCFMT12}. Linearizing~\eqref{eq:CCFMT12} about this equilibrium, we obtain
\begin{align}
\nonumber
\dot{v}_i(t) & = \, \beta_{i-1}^* v_{i-1}(t - \tau_{i-1}) - \beta_i^* v_{i}(t - \tau_{i}), \\ \label{eq:LCCFMTi}
\dot{y}_i(t) & = \, v_i(t),
\end{align}
for $i \in \lbrace 1, 2, \cdots, N \rbrace.$ Here, $\beta_i^* = \alpha_{i} (\dot{x}_0)^m / (b_i)^l$ denotes the equilibrium coefficient for the $i^{th}$ vehicle.

Notice from~\eqref{eq:LCCFMTi} that the evolution of $v_i(t),$ in the vicinity of its equilibrium, is not affected by the evolution of $y_i(t).$ Further, $y_i(t)$ can be obtained by integrating $v_i(t).$ Thus, we drop the variables $\lbrace y_i(t) \rbrace_{i = 1}^{N}$ when dealing with the linearized system. This yields
\begin{align}
\label{eq:LCCFMT}
\dot{v}_i(t) = \, \beta_{i-1}^* v_{i-1}(t - \tau_{i-1}) - \beta_i^* v_{i}(t - \tau_{i}).
\end{align}
In the remainder of this paper, we study system~\eqref{eq:LCCFMT} to deduce various conditions for the CCFM. It may be noted that~\eqref{eq:LCCFMT} is similar in form to the linearized RCCFM~\cite[Equation (3)]{GKK}. However, the equilibrium coefficient $\beta_i^*$ now accounts for the non-linearity parameter $l \in \mathbb{R}_+.$


%% file: sec3_ideal.tex
In this section, we consider the idealistic case of drivers that can react instantaneously to stimuli. This results in zero reactions delays, and hence the linear model described by system~\eqref{eq:LCCFMT} boils down to the following system of Ordinary Differential Equations (ODEs):
\begin{align}
\label{eq:LRCCFMTODE}
\dot{v}_i(t) = \, \beta_{i-1}^* v_{i-1}(t) - \beta_i^* v_{i}(t),
\end{align}
for $i \in \lbrace 1, 2, \cdots, N \rbrace.$ This can be succinctly written in matrix form as follows:
\begin{align}
\label{eq:LRCCFMTmatrix}
\dot{\textbf{V}}(t) = \, A \textbf{V}(t),
\end{align}
where $\textbf{V}(t) = [v_1(t) \, v_2(t) \, \cdots \, v_N(t)]^T \in \mathbb{R}^N,$ and $A \in \mathbb{R}^{N \times N}.$ The matrix $A,$ known as the \emph{dynamics matrix}~\cite[Section 2.2]{KJA}, is a lower-triangular matrix, given by:
\begin{align*}
A_{ij} = \begin{cases}
- \beta_i^*, & i = j, \\
\beta_j^*, & i = j+1, \\
0, & \text{elsewhere}.
\end{cases}
\end{align*}

To characterize the stability of system~\eqref{eq:LRCCFMTODE}, we require the eigenvalues of the dynamics matrix corresponding to system~\eqref{eq:LRCCFMTmatrix} to be negative~\cite[Theorem 5.1.1]{GL}. Since $A$ is a lower-triangular matrix, the characteristic polynomial is given by the product of the diagonal elements of the matrix $(\lambda I - A)$~\cite[Lemma 6.9.1]{INH}. Therefore, we have
\begin{align}
\label{eq:LRCCFMTchareq}
f(\lambda) = \text{det}(\lambda I - A) = \prod_{i = 1}^N \left(\lambda + \beta_i^*\right) = 0.
\end{align}
Therefore, eigenvalues corresponding to system~\eqref{eq:LRCCFMTODE} are located at $-\beta_i^*,$ $i \in \lbrace 1, 2, \cdots, N \rbrace.$ Note that, from physical constraints, $\alpha_i > 0$ and $b_i > 0$ $\forall i.$ This ensures $\beta_i^* > 0$ $\forall i,$ for all physically relevant systems. Hence, the corresponding eigenvalues will lie in the open left-half of the Argand plane, thereby ensuring the stability of system~\eqref{eq:LCCFMT} for all physically relevant values of the parameters.

%% file: sec4_small.tex
In this section, we analyze system~\eqref{eq:CCFMT12} in the small-delay regime. A way to obtain insights for small delays is to conduct a linearization on time. Thus, we obtain a system of ODEs, which serves as an approximation to the original infinite-dimensional system~\eqref{eq:LCCFMT}, for small delays. We derive the criterion for this system of ODEs to be stable, thereby emphasizing the design trade-off inherent among various system parameters and the reaction delay.


We begin by applying the Taylor series approximation to the time-delayed state variables thus: $ v_i(t-\tau_i) \approx v_i(t) - \tau_i \dot{v}_i(t). $ Using this approximation for terms in~\eqref{eq:LCCFMT}, and re-arranging the resulting equations, we obtain
\begin{align}
\label{eq:LRCCFMTsmall}
\dot{v}_i(t) + \frac{\beta_{i-1}^* \tau_{i-1}}{1- \beta_{i}^* \tau_{i}} \dot{v}_{i-1}(t) = \, \frac{\beta_{i-1}^*}{1- \beta_{i}^* \tau_{i}} v_{i-1}(t) - \frac{\beta_{i}^*}{1- \beta_{i}^* \tau_{i}} v_{i}(t),
\end{align}
for $i \in \lbrace 1, 2, \cdots, N \rbrace.$ This can be succinctly written in matrix form as
\begin{align}
\label{eq:LRCCFMTsmallmatrix}
B \dot{\textbf{V}}(t) = A_s \textbf{V}(t),
\end{align}
where $\textbf{V}(t) = [v_1(t) \, v_2(t) \, \cdots \, v_N(t)]^T \in \mathbb{R}^N.$ The matrix $A_s$ is as defined
\begin{align*}
A_{s_{ij}} = \begin{cases}
- \frac{\beta_i^*}{1-\beta_i^* \tau_i^*}, & i = j, \\
\frac{\beta_j^*}{1-\beta_i^* \tau_i^*}, & i = j+1, \\
0, & \text{elsewhere},
\end{cases}
\end{align*}
 and $B$ is given by
\begin{align*}
B_{ij} = \begin{cases}
1, & i = j, \\
\frac{-\beta_j^* \tau_j}{1 - \beta_i^* \tau_i}, & i = j + 1, \\
0, & \text{elsewhere}.
\end{cases}
\end{align*}
Note that $B$ is a lower-triangular matrix with unit diagonal entries. Hence, it is invertible, and the inverse is also a lower-triangular matrix having unit diagonal elements. Further, since $A$ is a lower-triangular matrix as well, the dynamics matrix corresponding to system~\eqref{eq:LRCCFMTsmall}, \emph{i.e.,} $\tilde{A} = B^{-1}A_s,$ is a lower-triangular matrix since it is the product of two lower-triangular matrices~\cite[Section 1.4]{GS}. Further, due to the said structures, the diagonal elements of $\tilde{A}$ are given by
\begin{align}
\label{eq:LRCCFMTsmalldiag}
\tilde{A}_{ii} = \frac{\beta_{i}^*}{1 - \beta_i^* \tau_i}, \text{ } i \geq 1.
\end{align}
Therefore, the characteristic polynomial corresponding to system~\eqref{eq:LRCCFMTsmall} is the product of the diagonal entries of the matrix $(\lambda I - \tilde{A})$~\cite[Lemma 6.9.1]{INH}. That is,
\begin{align}
\label{eq:LRCCFMTsmallchareq}
f(\lambda) = \text{det } (\lambda I - \tilde{A}) = \prod_{i = 1}^{N} \left(\lambda + \tilde{A}_{ii}\right) = 0.
\end{align}
This shows that the eigenvalues of system~\eqref{eq:LRCCFMTsmall} are located at $-\tilde{A}_{ii},$ $i \in \lbrace 1, 2, \cdots, N \rbrace.$  Hence, for system~\eqref{eq:LRCCFMTsmall} to be stable, the diagonal entries of its dynamics matrix $\tilde{A}$ have to be positive. From~\eqref{eq:LRCCFMTsmalldiag}, this is satisfied if and only if
\begin{align}
\label{eq:LRCCFMTsmallcond}
\beta_i^* \tau_i < 1, \text{ } i \in \lbrace 1, 2, \cdots, N \rbrace.
\end{align}
Hence, the above equation represents the necessary and sufficient condition for stability of the time-linearized system~\eqref{eq:LRCCFMTsmall}. Further, as noted in Section~\ref{sec:models},~\eqref{eq:LRCCFMTsmallcond} is a sufficient condition for the local stability of the CCFM, described by system~\eqref{eq:CCFMT12}.

%% file: sec5_hopf.tex
Having studied system~\eqref{eq:CCFMT12} in the no-delay and the small-delay regimes, in this section, we focus on the arbitrary-delay regime. We derive the necessary and sufficient condition for the local stability of system~\eqref{eq:CCFMT12}, and show that the corresponding traffic flow transits from the locally stable to the unstable regime via a Hopf bifurcation~\cite{BDH}.

\subsection{Transversality condition}
\label{section:RCCFMHBA}

Hopf bifurcation is a phenomenon wherein a system undergoes a stability switch due to a pair of conjugate eigenvalues crossing the imaginary axis in the Argand plane~\cite[Chapter 11, Theorem 1.1]{HL}. Mathematically, a Hopf bifurcation analysis is a rigorous way of proving the emergence of limit cycles in non-linear dynamical systems.

In order to ascertain whether the CCFM undergoes a stability loss via a Hopf bifurcation, we follow~\cite{GR} and introduce an exogenous, non-dimensional parameter $\kappa \, > \, 0.$ A general system of delay differential equations $\dot{x}(t) = f(x(t), x(t-\tau_1), \cdots, x(t-\tau_n))$ is modified to $\dot{x}(t) = \kappa f(x(t), x(t-\tau_1), \cdots, x(t-\tau_n))$ with the introduction of the exogenous parameter. In the specific case of the CCFM, introducing $\kappa$ in~\eqref{eq:CCFMT12} results in
\begin{align}
\nonumber
\dot{v}_i(t) & = \kappa \beta_{i-1}(t - \tau_{i-1}) v_{i-1}(t - \tau_{i-1}) - \kappa \beta_{i}(t - \tau_i) v_{i}(t - \tau_i), \\ \label{eq:modMOVMeqn}
\dot{y}_i(t) & = \kappa v_i(t),
\end{align}
for each $i \in \lbrace 1, 2, \cdots, N \rbrace.$ We linearize this about the all-zero equilibrium, and drop $y_i$'s, to obtain
\begin{align}
\label{eq:LRCCFMK}
\dot{v}_i(t) = \, \kappa \beta_{i-1}^* v_{i-1}(t - \tau_{i-1}) - \kappa \beta_i^* v_{i}(t - \tau_{i}),
\end{align}
for $i \in \lbrace 1, 2, \cdots, N \rbrace.$ The characteristic equation pertaining to~\eqref{eq:LRCCFMK} is~\cite[Equation (15)]{GKK}
\begin{align}
\label{eq:CELRCFFMK}
\lambda + \kappa \beta_i^* e^{- \lambda \tau_i} = \, 0.
\end{align}

It is well known that for~\eqref{eq:LRCCFMK} to be stable, all the roots of~\eqref{eq:CELRCFFMK} must lie in the open-left half of the Argand plane~\cite[Theorem 5.1.1]{GL}. Hence, to analyze the local stability of system~\eqref{eq:modMOVMeqn}, we search for a conjugate pair of eigenvalues of~\eqref{eq:CELRCFFMK} that crosses the imaginary axis in the Argand plane, thereby pushing the system into an unstable regime. To that end, we substitute $\lambda = j \omega$, with $j = \sqrt{-1}$, in~\eqref{eq:CELRCFFMK} to obtain
\begin{align*}
\kappa \beta_i^* \text{ cos}(\omega \tau_i) = 0, \text{ and } \omega - \kappa \beta_i^* \text{ sin}(\omega \tau_i) = 0.
\end{align*}
The first equality implies $\omega \tau_i$ = $(2n+1)\frac{\pi}{2}$ for $n = 0, 1, 2, \cdots$. Using this, the second equality then results in $\kappa \beta_i^*$ = $\omega$ for $n = 0, 2, 4, \cdots$. Therefore, when a conjugate pair of eigenvalues is on the imaginary axis in the Argand plane, we have
\begin{align}
\label{eq:RCCFMON}
\omega_0 = & \, (2n + 1) \frac{\pi}{2 \tau_i}, \text{ } n = 0, 1, 2, \cdots, \\ \label{eq:RCCFMKC}
\kappa_{cr} = & \, (2n + 1) \frac{\pi}{2 \beta_i^* \tau_i}, \text{ } n = 0, 2, 4, \cdots,
\end{align}
where $\kappa_{cr}$ is the critical value of $\kappa$ at $\omega = \omega_0$.

To show that system~\eqref{eq:modMOVMeqn} undergoes a Hopf bifurcation at $\kappa = \kappa_{cr}$ for each $n \in \lbrace 0, 2, 4, \cdots \rbrace$, we need to prove the transversality condition of the Hopf spectrum. That is, we must show that~\cite[Chapter 11, Theorem 1.1]{HL}
\begin{align}
\label{eq:HopfTransCond}
\text{Re}\left[ \frac{\text{d} \lambda}{\text{d} \kappa} \right]_{\kappa = \kappa_{cr} } \neq \, 0
\end{align}
holds for each $n \in \lbrace 0, 2, 4, \cdots \rbrace$. Therefore, we differentiate~\eqref{eq:CELRCFFMK} with respect to $\kappa$. Algebraic manipulations then yield
\begin{align}
\label{eq:RCCFMtranscond}
\text{Re}\left[ \frac{\text{d} \lambda}{\text{d} \kappa} \right]_{\kappa = \kappa_{cr} } = \frac{2 \beta^*_i \tau_i^2 \omega_0^2 }{(2n + 1)(1 + \tau_i^2 \omega_0^2)\pi} > 0,
\end{align}
for $n \in \lbrace 0, 2, 4, \cdots \rbrace$. This implies that system~\eqref{eq:modMOVMeqn} undergoes a Hopf bifurcation at $\kappa = \kappa_{cr}$ for each $n \in \lbrace 0, 2, 4, \cdots \rbrace$. Hence, $\kappa < \kappa_{cr}$ when $n = 0$ is the necessary and sufficient condition for system~\eqref{eq:modMOVMeqn} to be locally stable.

\begin{figure*}
 \begin{center}
 \subfloat[]{ 
\includegraphics[scale=0.23,angle=90]{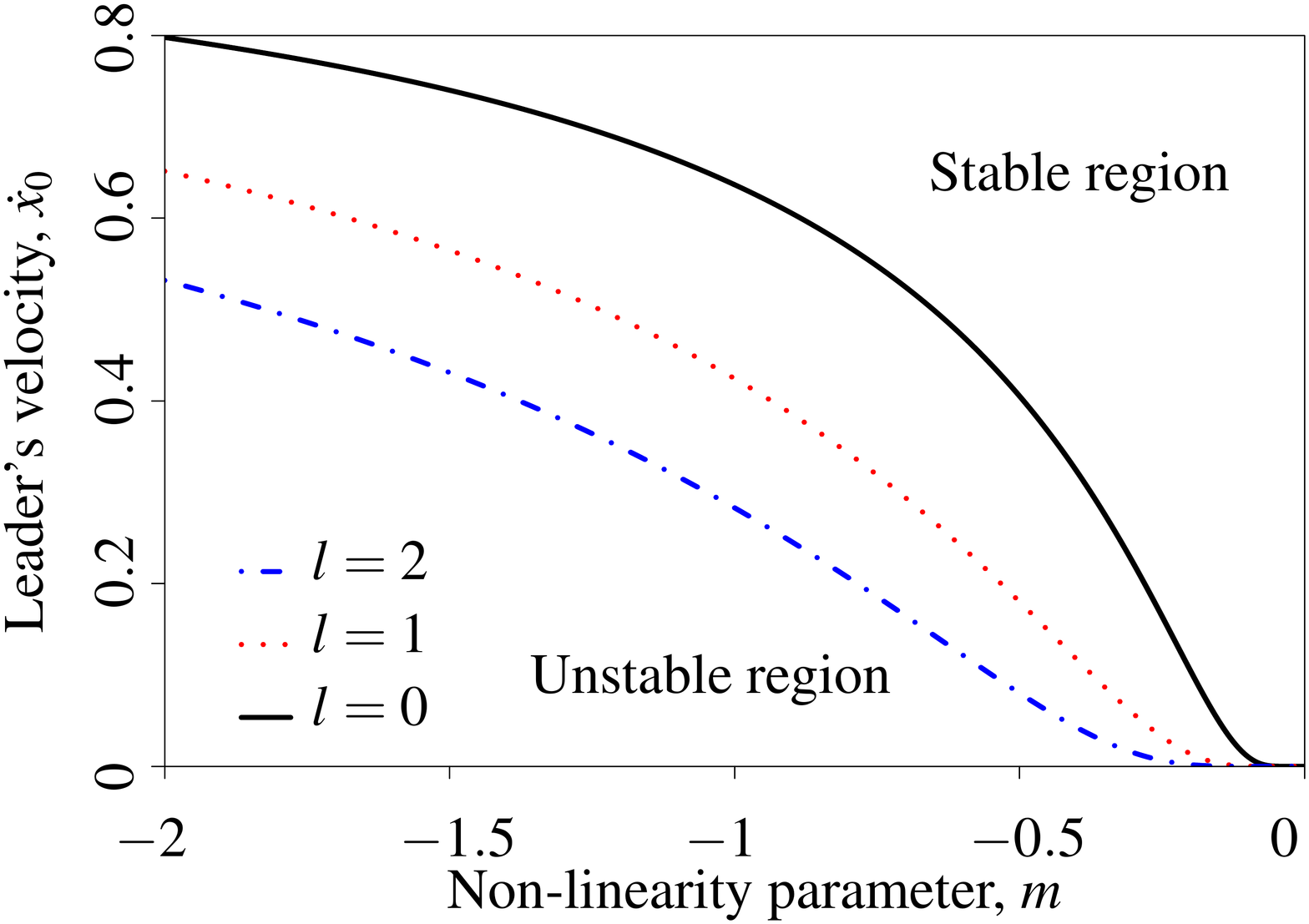}
  \label{fig:stability_chart_neg}
 } \hspace{20mm}
 \subfloat[]{
\includegraphics[scale=0.23,angle=90]{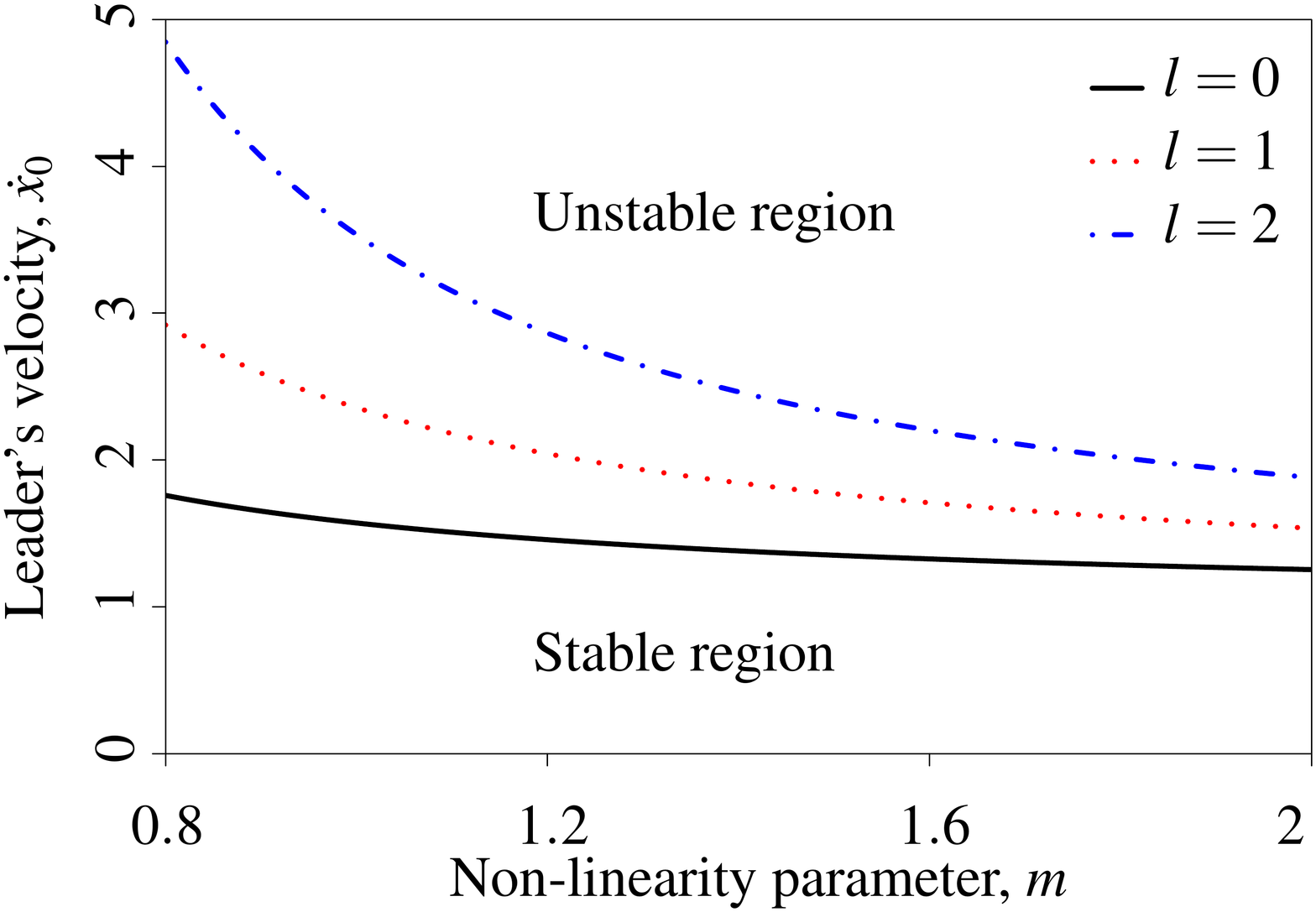}
\label{fig:stability_chart_pos}
  }
\caption{The local stability regions for the CCFM are depicted, with variation in non-linearity parameter $l$; a visual representation of~\eqref{eq:neccsuffRCCFMfin} with $c=1$ and $b_i > 1.$ (a) is for $m < 0$, whereas (b) is for $m > 0.$ Pictorially, we restrict $m \in [0.8,2]$ for clarity of visual representation. As $l$ increases, the CCFM becomes resilient to instability since $b_i >1.$}
\end{center}
\end{figure*}

First, we remark that $\kappa = \kappa_{cr}$ is the equation of the stability boundary, also known as the Hopf boundary. Once we obtain the expression for $\kappa_{cr},$ we tune the system parameters such that the non-dimensional parameter is unity on the stability boundary, $i.e.,$ we set $2 \beta_i^* \tau_i = \pi$ to make $\kappa_{cr}$ unity. Next, note that the system loses stability when the very first conjugate pair of eigenvalues, corresponding to $n = 0$ in~\eqref{eq:RCCFMKC}, crosses the imaginary axis. Further increase in $\kappa$ cannot restore system stability -- indeed, the derivative in~\eqref{eq:RCCFMtranscond} is positive for each $n \in \lbrace 0, 2, 4, \cdots \rbrace$. That is, an increase in $\kappa$ results in the eigenvalues moving to the right in the Argand plane, making it impossible to regain stability. Lastly, it is clear from~\eqref{eq:RCCFMKC} that $\alpha_i$ and $\tau_i$ are inversely related on the Hopf boundary, $i.e.,$ when $\kappa_{cr} = 1$. Hence, we set $\alpha_i \tau_i = c,$ a real constant, in order to study the trade-off between the leader's profile $\dot{x}_0$, and the non-linearity parameters $l$ and $m.$ The resulting necessary and sufficient condition for the local stability of system~\eqref{eq:CCFMT12} is
\begin{align}
\label{eq:neccsuffRCCFMfin}
\frac{( \dot{x}_0 )^m}{( b_i )^l} < \frac{\pi}{2c}.
\end{align}
Notice that we recover the necessary and sufficient condition for the local stability of the RCCFM~\cite[Equation (20)]{GKK} if $(i)$ the non-linearity parameter $l$ is set to zero, or $(ii)$ the equilibrium headway $b_i$ is set to unity. For these cases, the inference drawn in~\cite{GKK} holds: When $m > 0,$ slow lead vehicles stabilize the system, and for $m < 0,$ fast lead vehicles are required to ensure system stability. From Fig.s~\ref{fig:stability_chart_neg} and~\ref{fig:stability_chart_pos}, notice that the above inference holds for $l > 0$ as well. However, note that the non-linearity parameter $l$ affects the resilience of the CCFM to instability. Specifically, if the equilibrium headway $b_i > 1,$ then the locally stable region \emph{expands} with an increase in $l.$ However, when $b_i < 1,$ the locally stable region \emph{shrinks} with an increase in $l.$


\subsection{Discussion}
\label{section:RCCFMdisc}
A few comments are in order.
\begin{itemize}
\item[(1)] The foregoing analysis serves to clarify that the oscillations in state variables are a manifestation of \emph{limit cycles} (isolated closed orbits in phase space) that emerge due to a Hopf bifurcation, and not \emph{centers} (family of concentric closed orbits) as asserted in~\cite{XZ}. Further, as pointed out in the Introduction, these emergent limit cycles physically manifest themselves as a back-propagating congestion wave, known as a `phantom jam.' Therefore, the foregoing analysis offers a possible explanation of a commonly-observed phenomenon.
\item[(2)] Note that the non-dimensional parameter $\kappa$ introduced in Section \ref{section:RCCFMHBA} is not a system parameter; it is an exogenous mathematical entity to aid the analysis. Its usefulness is at the edge of the stability boundary, wherein it is used to push the system into the unstable regime in a controlled manner, as described in the analysis.
\item[(3)] It is well known in the control literature that a suitable variation in gain parameter can destabilize a system~\cite[Section 3.7]{RR}. Hence, to ensure that the bifurcation phenomenon is not an artifact of the exogenous parameter, it is required to verify that the transversality condition of the Hopf spectrum is satisfied for at least one system parameter beforehand. For the case of the CCFM, following the derivation in Section~\ref{section:RCCFMHBA}, it is easy to prove that the CCFM could undergo a Hopf bifurcation due to an appropriate variation in any of $\alpha_i,$ $\tau_i,$ $\dot{x}_0,$ $l$ or $m.$
\item[(4)] Note that the D-partitioning and its ``dual'' $\tau$-decomposition approaches~\cite[Section 3.3]{MS} are used extensively in the literature to study local stability properties of delay-differential equations. While the former assumes the delay to be fixed and independent of other parameters, the latter allows only the delay to be varied. In contrast, our approach allows stability analysis to be conducted by a continuous variation of \emph{any} parameter (including the exogenous parameter). Additionally, the use of an exogenous parameter as the bifurcation parameter captures any inter-dependence among model parameters, and generally simplifies the resulting algebra. Further, note that the bifurcation approach helps understand \emph{how} local stability is lost and also approximates the trajectory of the CCFM in the vicinity of the equilibrium using non-linear terms (up to third order in most cases) -- key additions in comparison to other widely-used approaches. This helps deduce the stability of the emergent limit cycles. The said analysis for the CCFM can be found in Section~\ref{section:RCCFMhopf}.
\item[(5)] Substituting $n = 0$ in~\eqref{eq:RCCFMKC}, and letting $\kappa = 1$ on the stability boundary, the necessary and sufficient condition for the local stability of system~\eqref{eq:CCFMT12} becomes
\begin{align}
\label{eq:RCCFMTstabcondsimplest}
\beta_i^* \tau_i < \frac{\pi}{2}.
\end{align}
Note that when $\tau_i = 0,$ \eqref{eq:RCCFMTstabcondsimplest} is trivially satisfied. This, in turn, implies that the CCFM is stable for all  parameter values, in the absence of reaction delays as seen in Section~\ref{sec:ideal}. However, as the delay increases,~\eqref{eq:RCCFMTstabcondsimplest} will be violated, thus resulting in loss of local stability of the CCFM. This, in turn, validates our claim that delays play an important role in determining the qualitative behavior of the CCFM. 
\item[(6)] Note that~\eqref{eq:RCCFMTstabcondsimplest} coincides with the necessary and sufficient condition derived in~\cite[Section 3.1]{XZ}. In fact, the characteristic equation of the form~\eqref{eq:CELRCFFMK} (with $\kappa = 1$) arise in several applications including population dynamics~\cite{KG}, engineering~\cite{HL}, consensus dynamics~\cite{OSM} and vehicular dynamics~\cite{XZ}. In general, such equations have been analyzed using both time-domain~\cite{GL, KG} and spectral-domain methods~\cite{MS, XZ}. However, to the best of our knowledge, none of these works apply the method used in this paper. Further, note that the evolution equations are non-linear time-delay equations. Hence, the analysis goes beyond that of a linear time-delay system; see Section~\ref{section:RCCFMhopf} for details.
\end{itemize}



%% file: sec6_noc.tex
In this section, we characterize the region of non-oscillatory convergence. Mathematically, this amounts to ensuring that the eigenvalues corresponding to system~\eqref{eq:LCCFMT} are negative real numbers. Qualitatively, non-oscillatory convergence avoids jerky vehicular motion since relative velocities and headways constitute dynamical variables. Such results could help ensure the smooth flow of traffic, and hence improve the ride quality.

In the above spirit, following~\cite{SD}, we derive the necessary and sufficient condition for non-oscillatory convergence of the CCFM. The characteristic equation pertaining to system~\eqref{eq:LCCFMT}, after dropping the subscript `$i$' for convenience, is $f(\lambda) = \lambda + \beta^* e^{-\lambda \tau} = 0$~\cite[Equation (8)]{GKK}.
Substituting $\lambda = - \sigma - j \omega$ and simplifying, we obtain
\begin{align}
\label{eq:RCCFMnoc1}
\sigma = \, \beta^* e^{\sigma \tau} \cos(\omega \tau), \text{ and } 
\omega = \, \beta^* e^{\sigma \tau} \sin(\omega \tau).
\end{align}
These, in turn, yield $\tan(\omega \tau) = \omega/\sigma.$ To ensure that $\omega = 0$ is the only solution of this equation, the necessary and sufficient condition is $\sigma \tau \geq 1.$ Re-writing~\eqref{eq:RCCFMnoc1}, we have
\begin{align*}
\beta^* \tau e^{\sigma \tau} \left( \frac{\sin(\omega \tau)}{\omega \tau} \right) = 1.
\end{align*}
In the limit $\omega \rightarrow 0,$ the term within the brackets represents $\text{sinc}(0) = 1.$ Moreover, the exponential term is bounded by $e$ since $\sigma \tau \geq 1.$ Hence, the boundary of non-oscillatory convergence is $\beta^* \tau e = 1,$ and the corresponding necessary and sufficient condition for non-oscillatory convergence is
\begin{align}
\label{eq:LRCCFMTnonoscil}
\beta^* \tau \leq \frac{1}{e}.
\end{align}

Notice that the region in the parameter space described by~\eqref{eq:LRCCFMTnonoscil} is a strict subset of the region described by~\eqref{eq:RCCFMTstabcondsimplest}. Therefore, from these two equations, we can summarize the conditions for the local stability of the CCFM as follows.
\begin{itemize}
\item[(1)] If $\beta^* \tau \in [0,\pi/2),$ the system is locally stable.
\item[(2)] Additionally, if $\beta^* \tau \in [0,1/e],$ the system converges asymptotically to the equilibrium in a non-oscillatory fashion.
\item[(3)] Contrarily, if $\beta^* \tau \in (1/e,\pi/2),$ the state variable oscillates about the equilibrium, converging asymptotically.
\end{itemize}

Note that, despite differing in the method of derivation,~\eqref{eq:LRCCFMTnonoscil} agrees with the condition for non-oscillatory condition derived in~\cite[Section 3]{XZ}.

%% file: sec7_roc.tex
Rate of convergence is an important performance metric that dictates the time a dynamical system takes to attain the desired equilibrium, when perturbed. In the context of a transportation network, it is related to the time required to attain the uniform traffic flow, once the traffic flow is perturbed (by events such as the departure of a vehicle from the platoon). Following~\cite{FB}, we characterize the rate of convergence for the CCFM.

The characteristic equation pertaining to system~\eqref{eq:LCCFMT}, with the subscript `$i$' dropped for ease of exposition, is $f(\lambda) = \lambda + \beta^* e^{-\lambda \tau} = 0$~\cite[Equation (8)]{GKK}. In time domain, this corresponds to a system $\dot{x}(t) = - \beta^* x(t - \tau),$ where $x$ is an arbitrarily chosen dynamical variable. The rate of convergence of such a system is the reciprocal of the smallest among $\sigma_1,$ $\sigma_2$ and $\sigma_3,$ where these quantities are obtained by solving the equations~\cite[Theorem 2]{FB}
\begin{align*}
\sigma \tau & = \, 1, \\
\sigma \tau e^{-\sigma \tau} & =  \beta^* \tau, \\
\frac{m}{\sin(m)} e^{-\frac{m}{\tan(m)}} = \beta^* \tau, & \hspace*{3mm} m = \sigma \tau \tan(m),
\end{align*}
respectively. The rate of convergence is maximum at $\tau^* = 1/ (\beta^* e).$ For $\tau < \tau^*,$ the rate of convergence increases, whereas it decreases for $\tau > \tau^*$~\cite{FB}.

We solve the above equations using MATLAB to illustrate the variations in the rate of convergence for the CCFM, as the reaction delay is varied. To that end, we consider a tagged vehicle following a lead vehicle with an equilibrium velocity of $10.$ The tagged vehicle has a sensitivity coefficient of $\alpha = 0.7$ and tries to maintain an equilibrium headway of $20.$ We fix $m = 2,$ and consider $l \in \lbrace 0.8, 1, 1.2 \rbrace .$

\begin{figure}[t]
\begin{center}
\includegraphics[scale=0.27,angle=90]{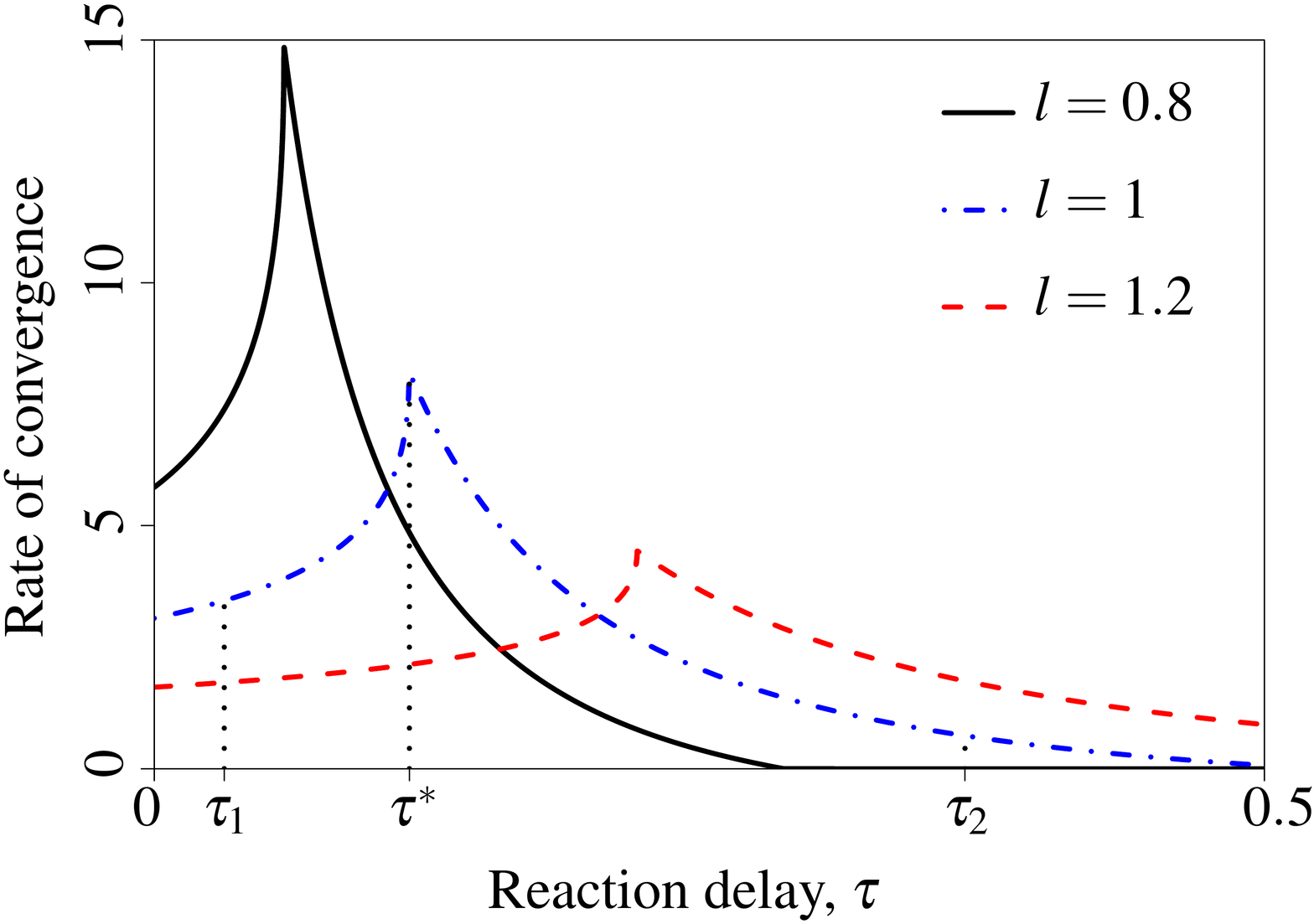}
\caption{Variation in the rate of convergence of the CCFM as the reaction delay is increased, for $l \in \lbrace 0.8, 1, 1.2 \rbrace.$}\label{fig:rocone}
\end{center}
\end{figure}

The rate of convergence for this system is plotted in Fig.~\ref{fig:rocone}. It can be seen that the rate of convergence increases with $\tau$ for $\tau < \tau^*,$ and decreases when the reaction delay is varied beyond $\tau^*.$ Also, note that the condition for the maximum rate of convergence coincides with the boundary for non-oscillatory convergence of the CCFM, $\beta^* \tau^* e = 1.$ Hence, it would be optimal to choose parameters satisfying this equation. The said figure portrays $\tau^*$ only for the $l = 1$ case.

However, in practice, system parameters may vary. This will result in a shift of the operating point of the CCFM, and may result in a trade-off between the rate of convergence and non-oscillatory convergence of the CCFM. Notice from Fig.~\ref{fig:rocone} that, for a given value of non-linearity parameter $l,$ the rate of convergence is not symmetric about $\tau^*.$ In the vicinity of $\tau^*,$ if the operating point of the CCFM shifts to the left of $\tau^*,$ the system retains its non-oscillatory behavior and the rate of convergence reduces drastically. On the other hand, if the operating point of the CCFM shifts to the right in the vicinity of $\tau^*$, the system converges to the equilibrium in an oscillatory fashion, but the reduction in the rate of convergence is not as drastic. However, if the reaction delay increases considerably beyond $\tau^*,$ then not only does the system exhibit oscillatory convergence, it may also converge to the equilibrium very slowly. This is portrayed in Fig.~\ref{fig:rocone} using $\tau_1 = \tau^*/3$ and $\tau_2 = 3 \tau^*,$ for $l = 1.$ Clearly, the rate of convergence at $\tau_2$ is much lesser than that at $\tau_1.$

Finally, notice from Fig.~\ref{fig:rocone} that, an increase in $l$ leads to a decrease in the rate of convergence. However, as discussed in Section~\ref{section:RCCFMHBA}, an increase in $l$ makes the system relatively resilient to instability since $b_1 = 20 > 1.$ Thus, there is a three-way trade-off involving the system's resilience to instability, rate of convergence and non-oscillatory convergence. Note that Fig.~\ref{fig:rocone} brings forth this trade-off for a fixed set of parameters. However, the same is true for other parameter values as well.

Note that the characteristic equation captures the closed-loop pairwise interaction in the platoon. To characterize the time taken by a platoon to reach an equilibrium (denoted by $T_{CCFM}^e$), we first define the time taken by the $i^{th}$ pair of vehicles in the platoon following the standard control-theoretic notion of ``settling time.'' That is, by $t_{i}^e(\epsilon),$ we denote the minimum time taken by the time-domain trajectory of the $i^{th}$ pair to enter, and subsequently remain within, the $\epsilon$-band around its equilibrium. For simplicity, we drop the explicit dependence on $\epsilon.$ Then, the platoon dynamics is said to converge to the uniform flow when the dynamics of each pair has settled inside the $\epsilon$-band of its respective equilibrium. Therefore, we have
\begin{align}
\label{eq:RCCFMequi}
T_{CCFM}^e = \max_{i = 1, 2, \dots, N} t_{i}^e.
\end{align}
Here, given $\epsilon > 0,$ $t_i^e$ is computed for the pair that has the least rate of convergence. This, in turn, yields $T_{CCFM}^e.$ Note that the convergence of the CCFM is asymptotic, \emph{i.e.,} the system does \emph{not} (strictly) converge to the equilibrium in finite time. Hence, we make use of the settling time concept.

%% file: sec8_hba.tex
In the previous sections, we have characterized the stable region for the CCFM, and studied two of its most important properties; namely, non-oscillatory convergence and the rate of convergence. We have also proved, by means of the transversality condition of the Hopf spectrum~\eqref{eq:RCCFMtranscond}, that system~\eqref{eq:CCFMT12} loses stability via a Hopf bifurcation. In this section, we study the CCFM when it is pushed just beyond the stable region. We characterize the \emph{type} of the bifurcation and the \emph{asymptotic orbital stability} of the emergent limit cycles, following closely the style of analysis presented in~\cite{BDH}, by using Poincar\'{e} normal forms and the center manifold theory.

We begin by denoting the non-linear part of the RHS of~\eqref{eq:modMOVMeqn} as $f_i.$ That is, for $i \in \lbrace 1, 2, \cdots, N \rbrace,$
\begin{align}
\label{eq:modelrhs}
f_i \triangleq \kappa \beta_{i-1}(t - \tau_{i-1}) v_{i-1}(t - \tau_{i-1}) - \kappa \beta_{i}(t - \tau_i) v_{i}(t - \tau_i).
\end{align}

Let $\mu = \kappa - \kappa_{cr}.$ Observe that the system undergoes a Hopf bifurcation at $\mu=0,$ where $\kappa = \kappa_{cr}.$ Henceforth, we consider $\mu$ as the bifurcation parameter. An incremental change in $\kappa$ from $\kappa_{cr}$ to $\kappa_{cr} + \mu,$ where $\mu>0,$ pushes the system in to its unstable regime. We now provide a concise step-by-step overview of the detailed local bifurcation analysis, before delving into the technical details.

\emph{Step 1}: Using Taylor series expansion, we segregate the RHS of~\eqref{eq:modelrhs} into linear and non-linear parts. We then cast this into the standard form of an Operator Differential Equation (OpDE).

\emph{Step 2}: At the critical value of the bifurcation parameter, \emph{i.e.,} at $\mu=0,$ the system has exactly one pair of purely imaginary eigenvalues with non-zero angular velocity, as given by~\eqref{eq:RCCFMON}. The linear eigenspace spanned by the corresponding eigenvectors is called the critical eigenspace. The center manifold theorem~\cite{BDH} guarantees the existence of a locally invariant $2$-dimensional manifold that is a tangent to the critical eigenspace at the equilibrium of the system.

\emph{Step 3}: Next, we project the system onto its critical eigenspace as well as its complement, at the critical value of the bifurcation parameter. This helps describe the dynamics of the system on the center manifold, with the aid of an ODE in a single complex variable.

\emph{Step 4}: Finally, using Poincar\'{e} normal forms, we evaluate the Lyapunov coefficient and the Floquet exponent, which characterize the type of the Hopf bifurcation and the asymptotic orbital stability of the emergent limit cycles respectively.

We begin the analysis by expanding~\eqref{eq:modelrhs} about the all-zero equilibrium using Taylor's series, to obtain
\begin{align}
\nonumber
\dot{v}_i(t) = & - \kappa \beta^*_i v_{i,t}( - \tau_i) + \kappa \beta^*_{i-1} v_{(i-1),t}(- \tau_{i-1}) - \frac{m}{\dot{x}_0} \beta_{i-1}^* v_{(i-1),t}^2(-\tau_{i-1}) + \frac{m}{\dot{x}_0} \beta_{i}^* v_{i,t}^2(-\tau_{i}) \\ \nonumber
& - \frac{m}{\dot{x}_0} \beta_{i-1}^* \sum\limits_{n = 1}^{i-2} v_{n,t}(-\tau_{i-1}) v_{(i-1),t}(- \tau_{i-1}) + \frac{m}{\dot{x}_0} \beta_{i-1}^* \sum\limits_{n = 1}^{i-1} v_{n,t}(-\tau_{i}) v_{i,t}(- \tau_{i}) \\ \nonumber
& - \frac{l}{b_{i-1}} \beta_{i-1}^* v_{(i-1),t}(-\tau_{i-1}) y_{(i-1),t}(-\tau_{i-1}) + \frac{l}{b_{i}} \beta_{i}^* v_{i,t}(-\tau_{i}) y_{i,t}(-\tau_{i}) 
\end{align}
\begin{align}
\nonumber
& + \frac{m(m-1)}{2 (\dot{x}_0)^{2}} \beta_{i-1}^* v_{i-1}^3(-\tau_{i-1}) - \frac{m(m-1)}{2 (\dot{x}_0)^{2}} \beta_i^* v_{i}^3(-\tau_{i}) - \frac{m(m-1)}{2 (\dot{x}_0)^{2}} \beta_i^* \sum\limits_{n = 1}^{i-1} \sum\limits_{k = 1}^{i-1} v_{i,t}(-\tau_{i}) v_{n,t}(-\tau_{i}) v_{k,t}(-\tau_{i}) \\ \nonumber
& + \frac{m(m-1)}{2 (\dot{x}_0)^{2}} \beta_{i-1}^* \sum\limits_{n = 1}^{i-2} \sum\limits_{k = 1}^{i-2} v_{(i-1),t}(-\tau_{i-1}) v_{n,t}(-\tau_{i-1}) v_{k,t}(-\tau_{i-1}) - \frac{2m(m-1)}{3(\dot{x}_0)^{2}} \beta_i^* \sum\limits_{n=1}^{i-1} v_{i,t}^2(-\tau_i) v_{n,t}(-\tau_i) \\ \nonumber
& + \frac{2m(m-1)}{3(\dot{x}_0)^{2}} \beta_{i-1}^* \sum\limits_{n=1}^{i-2} v_{(i-1),t}^*(-\tau_{i-1}) v_{n,t}(-\tau_{i-1}) - \frac{l m}{3(b_{i})(\dot{x}_0)} \beta_{i}^* \sum\limits_{n = 1}^{i-2} v_{i,t}(-\tau_{i}) v_{n,t}(-\tau_{i}) y_{i,t}(-\tau_{i}) \\ \nonumber
& + \frac{l m}{3(b_{i-1})(\dot{x}_0)} \beta_{i-1}^* \sum\limits_{n = 1}^{i-1} v_{(i-1),t}(-\tau_{i-1}) v_{n,t}(-\tau_{i-1}) y_{(i-1),t}(-\tau_{i-1}) - \frac{l m}{3(\dot{x}_0)(b_i)} \beta_i^* v_{i,t}^2(-\tau_i) y_{i,t}(-\tau_i) \\ \nonumber
& + \frac{l m}{3(\dot{x}_0)(b_{i-1})} \beta_{i-1}^* v_{(i-1),t}^2(-\tau_{i-1}) y_{(i-1),t}(-\tau_{i-1}) \\
\dot{y}_i(t) = & \kappa v_i(t), \label{eq:taylorexpanded}
\end{align}
where we use the shorthand $v_{i,t}(-\tau_i)$ to represent $v_i(t - \tau_i).$

In the following, we use $\mathcal{C}^k\left(A;B\right)$ to denote the linear space of all functions from $A$ to $B$ which are $k$ times differentiable, with each derivative being continuous. Also, we use $\mathcal{C}$ to denote $\mathcal{C}^0,$ for convenience. 

With the concatenated state $\textbf{S}(t),$ note that~\eqref{eq:CCFMT12} is of the form:
\begin{align}
\label{eq:tryandreduce}
\frac{\text{d}\textbf{S}(t)}{\text{d}t} = \mathcal{L}_{\mu} \textbf{S}_t(\theta) + \mathcal{F}(\textbf{S}_t(\theta), \mu),
\end{align}
where $t > 0$, $\mu \in \mathbb{R}$, and where for $\tau = \underset{i}{\text{max }} \tau_i > 0$,
\begin{align*}
\textbf{S}_t(\theta) = \textbf{S}(t + \theta), \text{ } \textbf{S}: [-\tau,0] \longrightarrow \mathbb{R}^{2N}, \text{ } \theta \in [-\tau, 0].
\end{align*}
Here, $\mathcal{L}_{\mu}: \mathcal{C}\left([-\tau,0];\mathbb{R}^{2N}\right) \longrightarrow \mathbb{R}^{2N}$ is a one-parameter family of continuous, bounded linear functionals, whereas the operator $\mathcal{F}: \mathcal{C}\left([-\tau,0];\mathbb{R}^{2N}\right) \longrightarrow \mathbb{R}^{2N}$ is an aggregation of the non-linear terms. Further, we assume that $\mathcal{F}(\textbf{S}_t, \mu)$ is analytic, and that $\mathcal{F}$ and $\mathcal{L}_{\mu}$ depend analytically on the bifurcation parameter $\mu$, for small $|\mu|$. The objective now is to cast~\eqref{eq:tryandreduce} in the standard form of an OpDE:
\begin{align}
\label{eq:reducetothis}
\frac{\text{d}\textbf{S}_t}{\text{d}t} = \mathcal{A}(\mu) \textbf{S}_t + \mathcal{R} \textbf{S}_t,
\end{align}
since the dependence here is on $\textbf{S}_t$ alone rather than both $\textbf{S}_t$ and $\textbf{S}(t)$. To that end, we begin by transforming the linear problem $\text{d}\textbf{S}(t) / \text{d}t = \mathcal{L}_{\mu} \textbf{S}_t(\theta)$. We note that, by the \emph{Riesz representation theorem}~\cite[Theorem 6.19]{WR}, there exists a $2N \times 2N$ matrix-valued measure $\eta(\cdot, \mu): \mathcal{B}\left(\mathcal{C}\left([-\tau,0];\mathbb{R}^{2N}\right)\right) \longrightarrow \mathbb{R}^{2N \times 2N}$, wherein each component of $\eta(\cdot)$ has bounded variation, and for all $\phi \in \mathcal{C}\left([-\tau,0];\mathbb{R}^{2N}\right),$ we have
\begin{align}
\label{eq:particularleqn}
\mathcal{L}_{\mu} \phi = \int\limits_{-\tau}^0 \text{d}\eta(\theta, \mu) \phi(\theta).
\end{align}
In particular,
\begin{align*}
\mathcal{L}_{\mu} \textbf{S}_t =  \int\limits_{-\tau}^0 \text{d}\eta(\theta, \mu) \textbf{S}(t +\theta).
\end{align*}

Motivated by the linearized system~\eqref{eq:LCCFMT}, we define
\begin{align*}
\text{d}\eta = 
\begin{bmatrix}
\tilde{A} & 0_{N \times N} \\
\kappa I_{N \times N} & 0_{N \times N}
\end{bmatrix} \text{d} \theta ,
\end{align*}
where \begin{equation*}
(\tilde{A})_{ij} = \begin{cases}
-\kappa \beta_i^* \delta(\theta + \tau_i), & i = j,\\
\kappa \beta_i^* \delta(\theta + \tau_i), & i = j + 1, j \geq 1,\\
0, & \text{otherwise.}
\end{cases}
\end{equation*}
For instance, when $N$ $=$ $2,$
\begin{equation*}
\text{d} \eta = 
\begin{bmatrix}
-\kappa \beta_1^* \delta(\theta + \tau_1) & 0 & 0 & 0 \\
\kappa \beta_1^* \delta(\theta + \tau_1) & - \kappa \beta_2^* \delta(\theta + \tau_2) & 0 & 0 \\
\kappa & 0 & 0 & 0 \\
0 & \kappa & 0 & 0
\end{bmatrix} \text{d} \theta.
\end{equation*}

For $\phi \in \mathcal{C}^1 \left([-\tau,0];\mathbb{C}^{2N}\right)$, we define
\begin{align}
\label{eq:aoperatordefinition}
\mathcal{A}(\mu) \phi(\theta) = \begin{cases}
\frac{\text{d} \phi(\theta)}{\text{d} \theta}, & \theta \in [-\tau,0),\\
\int\limits_{-\tau}^0 \text{d}\eta(s, \mu) \phi(s) \equiv \mathcal{L}_{\mu}, & \theta = 0,
\end{cases}
\end{align}
and
\begin{align*}
\mathcal{R} \phi(\theta) = \begin{cases}
0, & \theta \in [-\tau,0),\\
\mathcal{F}(\phi, \mu), & \theta = 0.
\end{cases}
\end{align*}

With the above definitions, we observe that $\text{d} \textbf{S}_t / \text{d} \theta \equiv \text{d} \textbf{S}_t / \text{d} t.$ Hence, we have successfully cast~\eqref{eq:tryandreduce} in the form of~\eqref{eq:reducetothis}. To obtain the required coefficients, it is sufficient to evaluate various expressions for $\mu = 0,$ which we use henceforth. We start by finding the eigenvector of the operator $\mathcal{A}(0)$ with eigenvalue $\lambda(0) = j \omega_0$. That is, we want an $2N \times 1$ vector (to be denoted by $q(\theta)$) with the property that $\mathcal{A}(0) q(\theta) = j \omega_0 q(\theta).$ We assume the form: $q(\theta) = [1 \text{ } \phi_1 \text{ } \phi_2 \cdots \text{ } \phi_{2N-1}]^T \text{ } e^{j \omega_0 \theta},$ and solve the eigenvalue equations. That is, we need to solve
\begin{equation*}
\begin{bmatrix}
-\kappa \beta_1^* e^{-j \omega_0 \tau_1} \\
\kappa \beta_1^* e^{-j \omega_0 \tau_1} - \kappa \beta_2^* e^{-j \omega_0 \tau_2} \\
\vdots \\
\kappa \beta_{N-1}^* e^{-j \omega_0 \tau_{N-1}} - \kappa \beta_N^* e^{-j \omega_0 \tau_{N}} \\
\kappa \Theta \\
\kappa \phi_1 \Theta \\
\vdots \\
\kappa \phi_{N-1} \Theta
\end{bmatrix} = j \omega_0
\begin{bmatrix}
1 \\
\phi_1 \\
\phi_2 \\
\vdots \\
\phi_{2N-2} \\
\phi_{2N-1}
\end{bmatrix},
\end{equation*}
where $\Theta = j ( e^{-j \omega_0 \tau} - 1) / \omega_0.$ This, in turn, necessitates the following assumption: $- \kappa \beta_1^* e^{- j \omega_0 \tau_1} = j \omega_0 \phi_0.$ Then, for $i \in \lbrace 1, 2, \cdots N-1 \rbrace,$ and $k \in \lbrace N, N+1, \cdots 2N-1 \rbrace,$
\begin{align*}
\phi_i = \frac{\kappa \beta_i^* e^{- j \omega_0 \tau_i} \phi_{i-1}}{j \omega_0 + \kappa \beta_{i+1}^* e^{- j \omega_0 \tau_{i+1}}}, \text{ and } \phi_k = \frac{\kappa \Theta  \phi_{N-k}}{j \omega_0},
\end{align*}
where we set $\phi_0 = 1$ for notational brevity.

We define the \emph{adjoint} operator as follows:
\begin{align*}
\mathcal{A}^*(0) \phi(\theta) = \begin{cases}
-\frac{\text{d} \phi(\theta)}{\text{d} \theta}, & \theta \in (0, \tau],\\
\int\limits_{-\tau}^0 \text{d}\eta^T(s, 0) \phi(-s), & \theta = 0,
\end{cases}
\end{align*}
where d$\eta^T$ is the transpose of d$\eta.$
We note that the domains of $\mathcal{A}$ and $\mathcal{A}^*$ are $\mathcal{C}^1\left([-\tau,0];\mathbb{C}^{2N}\right)$ and $\mathcal{C}^1\left([0,\tau];\mathbb{C}^{2N}\right)$ respectively. Therefore, if $j \omega_0$ is an eigenvalue of $\mathcal{A},$ then $-j \omega_0$ is an eigenvalue of $\mathcal{A}^*.$ Hence, to find the eigenvector of $\mathcal{A}^*(0)$ corresponding to $-j \omega_0,$ we assume the form: $p(\theta) = B [\psi_{2N-1} \text{ } \psi_{2N-2} \text{ } \psi_{2N-3} \text{ } \cdots \text{ } 1 ]^T \text{ } e^{j \omega_0 \theta},$ and solve $\mathcal{A}^*(0) p(\theta) = - j \omega_0 p(\theta).$ Simplifying this, we obtain
\begin{equation*}
\begin{bmatrix}
- \kappa \beta_1^* e^{j \omega_0 \tau_1} \psi_{2N-1} + \kappa \beta_1^* e^{j \omega_0 \tau_1} \psi_{2N-2} + \kappa \psi_{N-1} \tilde{\Theta} \\
- \kappa \beta_2^* e^{j \omega_0 \tau_2} \psi_{2N-2} + \kappa \beta_2^* e^{j \omega_0 \tau_2} \psi_{2N-3} + \kappa \psi_{N-2} \tilde{\Theta} \\
\vdots \\
- \kappa \beta_N^* e^{j \omega_0 \tau_N} \psi_N + \kappa \psi_0 \tilde{\Theta} \\
c \\
c \\
\vdots \\
c
\end{bmatrix} = - j \omega_0
\begin{bmatrix}
\psi_{2N-1} \\
\psi_{2N-2} \\
\psi_{2N-3} \\
\vdots \\
\psi_1 \\
1
\end{bmatrix}.
\end{equation*}
Here, we set $\psi_0 = 1$ for notational brevity and $c$ is some constant which can be shown to be $- j \omega_0 / B.$ Then, for $i \in \lbrace 1, 2, \cdots N-1 \rbrace,$ and $k \in \lbrace N+1, N+2, \cdots 2N-1 \rbrace,$ we have
\begin{align*}
\psi_N = \frac{\kappa \tilde{\Theta} \psi_0}{\kappa \beta_N^* e^{j \omega_0 \tau_{N}} - j \omega_0}, \text{ } \psi_{i} = - j \omega_0, \text{ and } \psi_i = \frac{\kappa \beta_{2N-i}^* e^{j \omega_0 \tau_{2N-i}} + \kappa \tilde{\Theta} \psi_{N-i}}{\kappa \beta_{2N-i}^* e^{j \omega_0 \tau_{2N-i}} - j \omega_0}.
\end{align*}

The normalization condition for Hopf bifurcation requires that $\langle p, q \rangle$ = $1,$ thus yielding an expression for $B.$

For any $q \in \mathcal{C}\left([-\tau,0];\mathbb{C}^{2N}\right)$ and $p \in \mathcal{C}\left([0, \tau];\mathbb{C}^{2N}\right)$, the inner product is defined as
\begin{align}
\label{eq:innerproductdefinition}
\langle p, q \rangle \triangleq \bar{p} \cdot q - \int\limits_{\theta = - \tau}^0 \int\limits_{\zeta = 0}^{\theta} \bar{p}^T(\zeta - \theta) \text{d}\eta q(\zeta) \text{ d}\zeta,
\end{align}
where the overbar represents the complex conjugate and the $``\cdot"$ represents the regular dot product. The value of $B$ such that the inner product between the eigenvectors of $\mathcal{A}$ and $\mathcal{A}^*$ is unity can be shown to be
\begin{align*}
B = \frac{1}{\zeta_1 + \zeta_2 + \zeta_3 + \zeta_4},
\end{align*}
where
\begin{align*}
\zeta_1 & = \overline{\Theta}^* \sum\limits_{i = 0}^{N-1} \kappa \psi_{N} \bar{\phi}_i, \text{ } \zeta_2 = \sum\limits_{i = 1}^{N-1} \kappa \beta_i^* \tau_i e^{j \omega_0 \tau_i} \overline{\phi}_{i-1} ( \psi_{2N-i} - \psi_{2N-i-1} ), \\
\zeta_3 & = - \kappa \beta_N^* \overline{\phi}_{N-1} \psi_N \tau_N e^{j \omega_0 \tau_N} , \text{ and } \zeta_4 = \sum\limits_{i = 0}^{2N-1} \psi_{2N-1-i} \bar{\phi}_i.
\end{align*}
In the above, we define $\phi_0 = \psi_0 = 0$ for notational brevity.

For $\textbf{S}_t$, a solution of~\eqref{eq:reducetothis} at $\mu$ $=$ $0,$ we define
\begin{align*}
z(t) = \langle p(\theta), \textbf{S}_t \rangle, \text{ and } \textbf{w}(t, \theta) = \textbf{S}_t(\theta) - 2 \text{Real} ( z(t) q(\theta) ).
\end{align*}
Then, on the center manifold $C_0$, we have $\textbf{w}(t, \theta)$ = $\textbf{w}(z(t), \bar{z}(t), \theta)$, where
\begin{align}
\label{eq:onthemanifold}
\textbf{w}(z(t), \bar{z}(t), \theta) = \textbf{w}_{20}(\theta) \frac{z^2}{2} + \textbf{w}_{02}(\theta) \frac{\bar{z}^2}{2} + \textbf{w}_{11}(\theta) z \bar{z} + \cdots .
\end{align}
Effectively, $z$ and $\bar{z}$ are the local coordinates for $C_0$ in $\mathcal{C}$ in the directions of $p$ and $\bar{p}$ respectively. We note that $\textbf{w}$ is real if $\textbf{S}_t$ is real, and we deal only with real solutions. The existence of the center manifold $C_0$ enables the reduction of~\eqref{eq:reducetothis} to an ODE in a single complex variable on $C_0$. At $\mu$ = 0, the said ODE can be described as
\begin{align}
\nonumber
\dot{z}(t) = & \, \left\langle p, \mathcal{A} \textbf{S}_t + \mathcal{R} \textbf{S}_t \right\rangle, \\ \nonumber
= & \, j \omega_0 z(t) + \bar{p}(0). \mathcal{F}\left( \textbf{w}(z, \bar{z}, \theta) + 2 \text{Real}( z(t) q(\theta) ) \right), \\
= & \, j \omega_0 z(t) + \bar{p}(0). \mathcal{F}_0(z, \bar{z}) \label{eq:odeforz}.
\end{align}
This is written in abbreviated form as
\begin{align}
\label{eq:odeforz2}
\dot{z}(t) = j \omega_0 z(t) + g(z, \bar{z}).
\end{align}
The objective now is to expand $g$ in powers of $z$ and $\bar{z}$. However, this requires $\textbf{w}_{ij}(\theta)$'s from~\eqref{eq:onthemanifold}. Once these are evaluated, the ODE~\eqref{eq:odeforz} for $z$ would be explicit (as given by~\eqref{eq:odeforz2}), where $g$ can be expanded in terms of $z$ and $\bar{z}$ as
\begin{align}
 \label{eq:gexpanded}
g(z, \bar{z}) = \, \bar{p}(0). \mathcal{F}_0(z, \bar{z}) = \, g_{20} \frac{z^2}{2} + g_{02} \frac{\bar{z}^2}{2} + g_{11} z \bar{z} + g_{21} \frac{z^2 \bar{z}}{2} + \cdots .
\end{align}
Next, we write $\dot{\textbf{w}} = \dot{\textbf{S}}_t - \dot{z} q - \dot{\bar{z}} \bar{q}.$ Using~\eqref{eq:reducetothis} and~\eqref{eq:odeforz2}, we then obtain the following ODE:
\begin{align*}
\dot{\textbf{w}} = \begin{cases}
\mathcal{A} \textbf{w} - 2 \text{Real} ( \bar{p}(0). \mathcal{F}_0 q(\theta) ), & \theta \in [-\tau, 0),\\
\mathcal{A} \textbf{w} - 2 \text{Real} ( \bar{p}(0). \mathcal{F}_0 q(0) ) + \mathcal{F}_0, & \theta = 0.
\end{cases}
\end{align*}
This can be re-written using \eqref{eq:onthemanifold} as
\begin{align}
\label{eq:rewrittenwode}
\dot{\textbf{w}} = \mathcal{A} \textbf{w} + H(z, \bar{z}, \theta),
\end{align}
where $H$ can be expanded as
\begin{align}
\label{eq:Hexpansion}
H(z, \bar{z}, \theta) = \, H_{20}(\theta) \frac{z^2}{2} + H_{02}(\theta) \frac{\bar{z}^2}{2} + H_{11}(\theta) z \bar{z} + H_{21}(\theta) \frac{z^2 \bar{z}}{2} + \cdots .
\end{align} 
Near the origin, on the manifold $C_0$, we have $\dot{\textbf{w}} = \textbf{w}_z \dot{z} + \textbf{w}_{\bar{z}} \dot{\bar{z}}.$ Using~\eqref{eq:onthemanifold} and~\eqref{eq:odeforz2} to replace $\textbf{w}_z \dot{z}$ (and their conjugates, by their power series expansion) and equating with~\eqref{eq:rewrittenwode}, we obtain the following operator equations:
\begin{align}
\label{eq:opeqn1}
(2 j \omega_0 - \mathcal{A}) \textbf{w}_{20}(\theta) = & \, H_{20}(\theta), \\ \label{eq:opeqn2}
- \mathcal{A} \textbf{w}_{11} = & \, H_{11}(\theta), \\ \label{eq:opeqn3}
-(2 j \omega_0 + \mathcal{A}) \textbf{w}_{02}(\theta) = & \, H_{02}(\theta).
\end{align}
We start by observing that
\begin{align*}
\textbf{S}_t(\theta) = \, \textbf{w}_{20}(\theta) \frac{z^2}{2} + \textbf{w}_{02}(\theta) \frac{\bar{z}^2}{2} + \textbf{w}_{11}(\theta) z \bar{z} \, + z q(\theta) + \bar{z} \bar{q}(\theta) + \cdots .
\end{align*}

From the Hopf bifurcation analysis~\cite{BDH}, we know that the coefficients of $z^2$, $\bar{z}^2$, $z^2 \bar{z}$, and $z \bar{z}$ terms are used to approximate the system dynamics. Hence, we only retain these terms in the expansions. 

To obtain the effect of non-linearities, we substitute the aforementioned terms appropriately in the non-linear terms of~\eqref{eq:taylorexpanded} and separate the terms as required. Therefore, for each $i \in \lbrace 1, 2, \dots, 2N \rbrace$, we have the non-linearity term to be
\begin{align}
\label{eq:nonlinvector}
\mathcal{F}_i  =  \mathcal{F}_{20i} \frac{z^2}{2} + \mathcal{F}_{02i} \frac{\bar{z}^2}{2} + \mathcal{F}_{11i} z \bar{z} + \mathcal{F}_{21i} \frac{z^2 \bar{z}}{2},
\end{align}
where, for $i \in \lbrace 1, 2, \cdots, N \rbrace,$ the coefficients are given by
\begin{align*}
\mathcal{F}_{20i} = & \beta_{i-1}^* w_{20(i-1)}(-\tau_{i-1}) - \beta_i^* w_{20i}(-\tau_i) - 4 \left( \frac{m}{\dot{x}_0} + \frac{l}{b_{i-1}} \right) \beta_{i-1}^* e^{- 2 j \omega_0 \tau_{i-1}} + 4 \left( \frac{m}{\dot{x}_0} + \frac{l}{b_{i}} \right) \beta_{i}^* e^{- 2 j \omega_0 \tau_{i}}, \\
& - 4 \left( \frac{m}{\dot{x}_0} \right) \beta_{i-1}^* (i-2) e^{-2 j \omega_0 \tau_{i-1}} + 4 \left( \frac{m}{\dot{x}_0} \right) \beta_{i}^* (i-1) e^{-2 j \omega_0 \tau_{i}}, \\
\mathcal{F}_{02i}  = & \beta_{i-1}^* w_{02(i-1)}(-\tau_{i-1}) - \beta_i^* w_{02i}(-\tau_i) - 4 \left( \frac{m}{\dot{x}_0} + \frac{l}{b_{i-1}} \right) \beta_{i-1}^* e^{2 j \omega_0 \tau_{i-1}} + 4 \left( \frac{m}{\dot{x}_0} + \frac{l}{b_{i}} \right) \beta_{i}^* e^{2 j \omega_0 \tau_{i}}, \\
& - 4 \left( \frac{m}{\dot{x}_0} \right) \beta_{i-1}^* (i-2) e^{2 j \omega_0 \tau_{i-1}} + 4 \left( \frac{m}{\dot{x}_0} \right) \beta_{i}^* (i-1) e^{2 j \omega_0 \tau_{i}}, \\
\mathcal{F}_{11i} = & \beta_{i-1}^* w_{11(i-1)}(-\tau_{i-1}) - \beta_{i}^* w_{11i}(-\tau_{i}) - 2 \left( \frac{m}{\dot{x}_0} + \frac{l}{b_{i-1}} \right) \beta_{i-1}^* + 2 \left( \frac{m}{\dot{x}_0} + \frac{l}{b_{i}} \right) \beta_{i}^*, \\
\mathcal{F}_{21i}  = & -2 \left( \frac{m}{\dot{x}_0} + \frac{l}{b_{i-1}} \right) \beta_{i-1}^* \left( w_{20(i-1)} e^{j \omega_0 \tau_{i-1}} + 2 w_{11(i-1)}(-\tau_{i-1}) e^{-j \omega_0 \tau_{i-1}} \right) \\
& + 2 \left( \frac{m}{\dot{x}_0} + \frac{l}{b_{i}} \right) \beta_{i}^* \left( w_{20i} e^{j \omega_0 \tau_{i}} + 2 w_{11i}(-\tau_{i}) e^{-j \omega_0 \tau_{i}} \right)
\end{align*}
\begin{align*}
& - \left(\frac{m}{\dot{x}_0}\right) \beta_{i-1}^* \sum\limits_{n = 1}^{i-2} \left[ (w_{20n}(-\tau_{i-1}) + w_{20(i-1)}(-\tau_{i-1})) e^{j \omega_0 \tau_{i-1}} + 2 (w_{11n}(-\tau_{i-1}) + w_{11(i-1)}(-\tau_{i-1})) e^{ -j \omega_0 \tau_{i-1}}  \right] \\
& + \left(\frac{m}{\dot{x}_0}\right) \beta_{i}^* \sum\limits_{n = 1}^{i-1} \left[ (w_{20n}(-\tau_{i}) + w_{20i}(-\tau_{i})) e^{j \omega_0 \tau_{i}} + 2 (w_{11n}(-\tau_{i}) + w_{11i}(-\tau_{i})) e^{ -j \omega_0 \tau_{i}}  \right] \\
& + (2 e^{- \omega_0 \tau_{i-1}} \beta_{i-1}^*) \left( \frac{m(m-1)}{2(\dot{x}_0)^2} + \frac{m(m-1) (i-2)^2}{(\dot{x}_0)^2} + \frac{2m(m-1)(i-2)}{3(\dot{x}_0)^2} + \frac{l m (i-2)}{3(b_{i-1}) (\dot{x}_0)} + \frac{l m}{3(b_{i-1}) (\dot{x}_0)} \right) \\
& - (2 e^{- \omega_0 \tau_{i}} \beta_{i}^*) \left( \frac{m(m-1)}{2(\dot{x}_0)^2} + \frac{m(m-1) (i-1)^2}{(\dot{x}_0)^2} + \frac{2m(m-1)(i-1)}{3(\dot{x}_0)^2} + \frac{l m (i-1)}{3(b_{i}) (\dot{x}_0)} + \frac{l m}{3(b_{i}) (\dot{x}_0)} \right).
\end{align*}

We represent the vector of non-linearities used in~\eqref{eq:odeforz} as $\mathcal{F}_0 = [\mathcal{F}_1 \text{ } \mathcal{F}_2 \text{ }  \text{ } \cdots \text{ } \mathcal{F}_N]^T.$ Next, we compute $g$ using $\mathcal{F}_0$ as
\begin{align}
 \label{eq:gequation}
g(z, \bar{z}) =  \bar{p}(0).\mathcal{F}_0 = \bar{B} \sum\limits_{l = 1}^N \bar{\psi}_{N-l} \mathcal{F}_l.
\end{align}
Substituting~\eqref{eq:nonlinvector} in~\eqref{eq:gequation}, and comparing with~\eqref{eq:gexpanded}, we obtain
\begin{align}
\label{eq:gxeq}
g_{x} = \bar{B} \sum\limits_{l = 1}^N \bar{\psi}_{N-l} \mathcal{F}_{xl},
\end{align}
where $x \in \lbrace 20, 02, 11, 21 \rbrace.$ Using~\eqref{eq:gxeq}, the corresponding coefficients can be computed. However, computing $g_{21}$ requires $\textbf{w}_{20}(\theta)$ and $\textbf{w}_{11}(\theta).$ Hence, we perform the requisite computation next. For $\theta$ $\in$ $[-\tau,0)$, $H$ can be simplified as
\begin{align*}
H(z, \bar{z}, \theta) & = - \text{Real} \left( \bar{p}(0). \mathcal{F}_0 q(\theta) \right), \\
& = - \left( g_{20} \frac{z^2}{2} + g_{02} \frac{\bar{z}^2}{2} + g_{11} z \bar{z} + \cdots \right) q(\theta) \\
& \hspace*{4mm} - \left( \bar{g}_{20} \frac{\bar{z}^2}{2} + \bar{g}_{02} \frac{z^2}{2} + \bar{g}_{11} z \bar{z} + \cdots \right) \bar{q}(\theta),
\end{align*}
which, when compared with~\eqref{eq:Hexpansion}, yields
\begin{align}
\label{eq:H20eqn}
H_{20}(\theta) & = - g_{20} q(\theta) - \bar{g}_{20} \bar{q}(\theta), \\ \label{eq:H11eqn}
H_{11}(\theta) & = - g_{11} q(\theta) - \bar{g}_{11} \bar{q}(\theta).
\end{align}
From~\eqref{eq:aoperatordefinition},~\eqref{eq:opeqn1} and~\eqref{eq:opeqn2}, we obtain the following ODEs:
\begin{align}
\label{eq:finode1}
\dot{\textbf{w}}_{20}(\theta) = & \, 2 j \omega_0 \textbf{w}_{20}(\theta) + g_{20} q(\theta) + \bar{g}_{02} \bar{q}(\theta), \\ \label{eq:finode2}
\dot{\textbf{w}}_{11}(\theta) = & \, g_{11} q(\theta) + \bar{g}_{11} \bar{q}(\theta).
\end{align}
Solving~\eqref{eq:finode1} and~\eqref{eq:finode2}, we obtain
\begin{align}
\label{eq:w20odesoln}
\textbf{w}_{20}(\theta) = & \, - \frac{g_{20}}{j \omega_0} q(0) e^{j \omega_0 \theta} - \frac{\bar{g}_{02}}{3 j \omega_0} \bar{q}(0) e^{-j \omega_0 \theta} + \textbf{e} \text{ } e^{2 j \omega \theta}, \\ \label{eq:w11odesoln}
\textbf{w}_{11}(\theta) = & \, \frac{g_{11}}{j \omega_0} q(0) e^{j \omega_0 \theta} - \frac{\bar{g}_{11}}{ j \omega_0} \bar{q}(0) e^{-j \omega_0 \theta} + \textbf{f},
\end{align}
for some vectors $\textbf{e}$ and $\textbf{f},$ to be determined.

To that end, we begin by defining the following vector: $\tilde{\mathcal{F}}_{20} \triangleq [\mathcal{F}_{201} \text{ } \mathcal{F}_{202} \text{ } \cdots \text{ } \mathcal{F}_{20N}]^T.$
Equating~\eqref{eq:opeqn1} and~\eqref{eq:H20eqn}, and simplifying, yields the operator equation: $2 j \omega_0 \textbf{e} - \mathcal{A} \left( \textbf{e} \text{ } e^{2 j \omega_0 \theta} \right) = \tilde{\mathcal{F}}_{20}.$ On simplification, we obtain
\begin{align*}
\begin{bmatrix}
\big(2 j \omega_0 + \kappa \beta_1^* \big) \textbf{e}_1 \\
\big(2 j \omega_0 + \kappa \beta_2^*  \big) \textbf{e}_2 - \kappa \beta_1^* \textbf{e}_{1} \\
\vdots \\
\big(2 j \omega_0 + \kappa \beta_N^* \big) \textbf{e}_N - \kappa \beta_{N-1}^* \textbf{e}_{N-1} \\
-\kappa \tau \textbf{e}_1 + 2 j \omega_0 \textbf{e}_{N+1} \\
\vdots \\
-\kappa \tau \textbf{e}_N + 2 j \omega_0 \textbf{e}_{2N} \\
\end{bmatrix}
= \tilde{\mathcal{F}}_{20}.
\end{align*}
This, in turn, yields
\begin{align}
\label{eq:evector}
\textbf{e}_i = \frac{\mathcal{F}_{20i} - \kappa \beta_{i-1}^* \textbf{e}_{i-1}}{2 j \omega_0 + \kappa \beta_i^* }, \text{ and, } \textbf{e}_{N+i} = \frac{\mathcal{F}_{20(N+i)} + \kappa \tau \textbf{e}_i}{2 j \omega_0},
\end{align}
for $i \in \lbrace 1, 2, \cdots, N \rbrace.$ Here, we set $\textbf{e}_0 = 0$ for notational brevity.

Next, equating~\eqref{eq:opeqn2} and~\eqref{eq:H11eqn}, and simplifying, we obtain the operator equation $\mathcal{A} \textbf{f} = -\tilde{\mathcal{F}}_{11},$ with $\tilde{\mathcal{F}}_{11} \triangleq [\mathcal{F}_{111} \text{ } \mathcal{F}_{112} \text{ } \cdots \text{ } \mathcal{F}_{11N}]^T.$ To solve this, we make the assumption that, for $i \in \lbrace 1, 2, \cdots, N-1 \rbrace,$
\begin{align*}
\mathcal{F}_{20i} \tau + \kappa \beta_{i-1}^* \textbf{f}_{i-1} + \beta_i^* \mathcal{F}_{11(N+i)} = 0
\end{align*}
Therefore, on simplification, the above-mentioned operator equation yields
\begin{align*}
\begin{bmatrix}
- \kappa \beta_1^* \textbf{f}_1 \\
\kappa \beta_1^* \textbf{f}_1 - \kappa \beta_2^* \textbf{f}_2 \\
\vdots \\
\kappa \beta_{N-1}^* \textbf{f}_{N-1} - \kappa \beta_N^* \textbf{f}_N \\
\kappa \tau \textbf{f}_1 \\
\vdots \\
\kappa \tau \textbf{f}_N \\ 
\end{bmatrix}
= - \tilde{\mathcal{F}}_{11}.
\end{align*}
On solving this, we obtain for $i \in \lbrace 1, 2, \cdots, N \rbrace,$
\begin{align}
\label{eq:fvector}
\textbf{f}_i = \frac{\mathcal{F}_{11i} + \kappa \beta_{i-1}^* \textbf{f}_{i-1}}{\kappa \beta_i^*}, \text{ and, } \textbf{f}_{N+i} = c,
\end{align}
where $c$ is an arbitrary constant, which we set to zero for simplicity. We also set $\textbf{f}_0 = 0$ for notational brevity.

Substituting for $\textbf{e}$ and $\textbf{f}$ from~\eqref{eq:evector} and~\eqref{eq:fvector} in~\eqref{eq:w20odesoln} and~\eqref{eq:w11odesoln} respectively, we obtain $\textbf{w}_{20}(\theta)$ and $\textbf{w}_{11}(\theta).$ This, in turn, facilitates the computation of $g_{21}.$ We can then compute
\begin{align*}
 c_1(0)  =  \frac{j}{2 \omega_0} \left( g_{20} g_{11} - 2 | g_{11} |^2 - \frac{1}{3} | g_{02} |^2 \right) + & \frac{g_{21}}{2}, \\
 \alpha^{'}(0) = \text{Re}\left[ \frac{\text{d} \lambda}{\text{d} \kappa} \right]_{\kappa = \kappa_{cr} },  \text{ } \mu_2 = - \frac{\text{Re}[c_1(0)]}{\alpha^{'}(0)}, \text{ and } & \beta_2 = 2 \text{Re}[c_1(0)].
\end{align*}

Here, $c_1(0)$ is known as the Lyapunov coefficient and $\beta_2$ is the Floquet exponent. It is known from~\cite{BDH} that these quantities are useful since
\begin{enumerate}
\item[$(i)$] If $\mu_2 > 0$, then the bifurcation is \emph{supercritical}, whereas if $\mu_2 < 0$, then the bifurcation is \emph{subcritical}.
\item[$(ii)$] If $\beta_2 > 0$, then the limit cycle is \emph{asymptotically orbitally unstable}, whereas if $\beta_2 < 0$, then the limit cycle is \emph{asymptotically orbitally stable}.
\end{enumerate}

We now present numerically-constructed bifurcation diagrams to gain some insight into the effect of various parameters on the amplitude of the limit cycle.

\begin{figure}[t]
\begin{center}
\includegraphics[scale=0.27,angle=90]{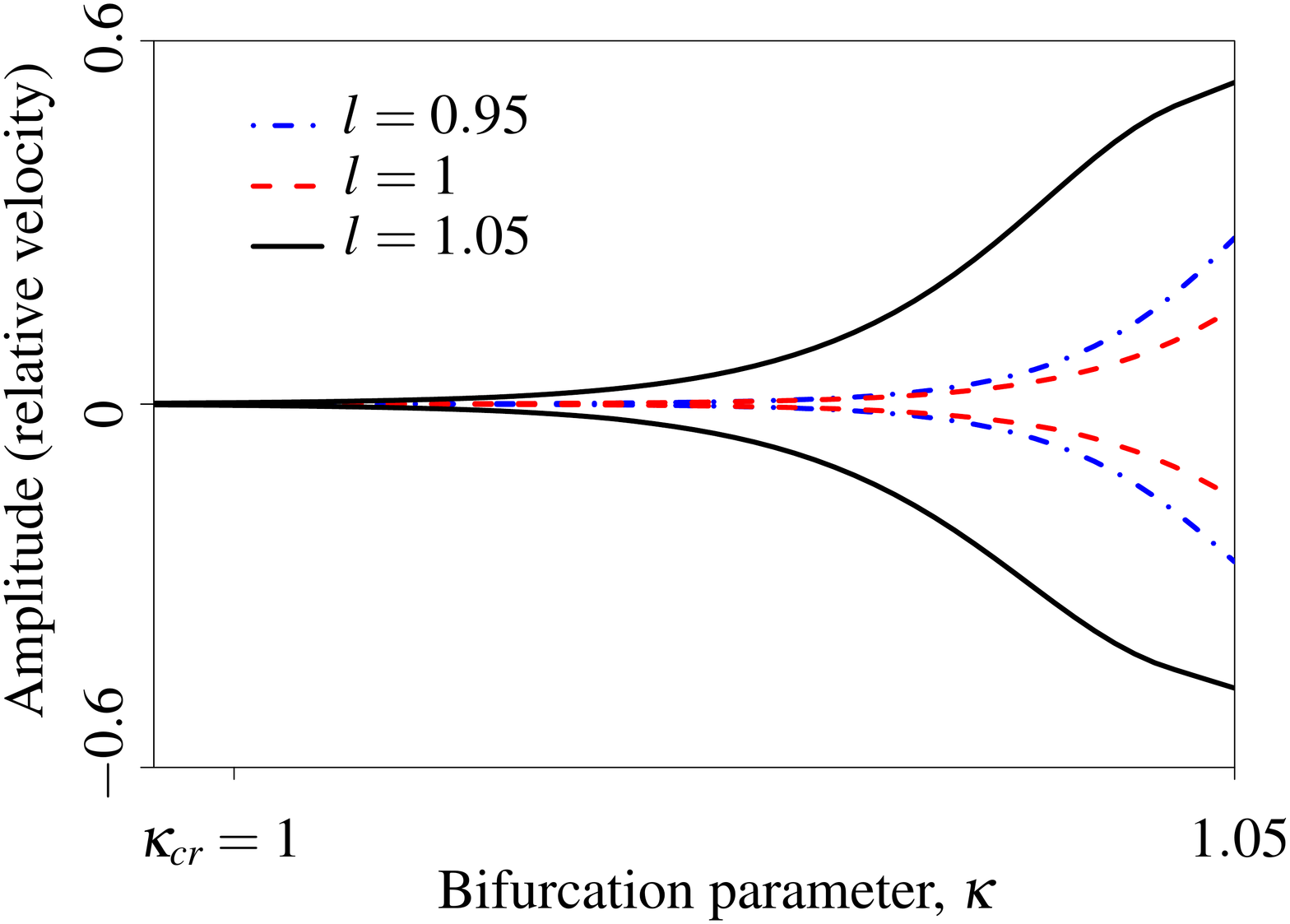}
\caption{\emph{Bifurcation diagram:} Variation in the amplitude of relative velocity of the CCFM as the non-dimensional parameter $\kappa$ is varied, for $l \in \lbrace 0.95, 1, 1.05 \rbrace.$}\label{fig:bif_diag}
\end{center}
\end{figure}

\subsection*{Bifurcation diagrams}

We next present bifurcation diagrams, numerically constructed using the scientific computation software MATLAB. We implement a discrete version of system~\eqref{eq:modMOVMeqn} with update time $T_s = 0.01$ s. We then vary the non-dimensional parameter $\kappa$ in the range $[1,1.05]$, and record the amplitude of the relative velocity in steady state. The resulting plot of the envelope of the relative velocity as a function of the non-dimensional parameter is called a \emph{bifurcation diagram}.

For illustration, we consider a single vehicle following a lead vehicle whose equilibrium velocity is $10.$ For the follower vehicle, we initialize the parameters as follows. $\alpha_1 = 0.7,$ $b_1 = 20$ and $m = 2.$ We set the reaction delay $\tau_1 = \tau_{cr} \approx 0.45,$ to ensure $\kappa_{cr} = 1.$ Next, we vary the non-dimensional parameter in the vicinity of unity, and record the resulting amplitude of the relative velocity for $l = 0.95, 1, 1.05.$ The resulting bifurcation diagram is portrayed in Fig.~\ref{fig:bif_diag}.

It can be inferred from Fig.~\ref{fig:bif_diag} that, there is no monotonicity in the amplitude of relative velocity with an increase in the non-linearity parameter $l.$ This is unlike the result presented in~\cite{GKK}, wherein monotonicity of the amplitude of relative velocity with an increase in $m$ was shown numerically, for $l = 0$ (the RCCFM). While Fig.~\ref{fig:bif_diag} was constructed for $m = 2,$ extensive computations reveal a lack of monotonicity in the amplitude of limit cycle with an increase in $l.$

%% file: sec9_sims.tex
\begin{figure}[h]
\begin{center}
\includegraphics[scale=0.27,angle=90]{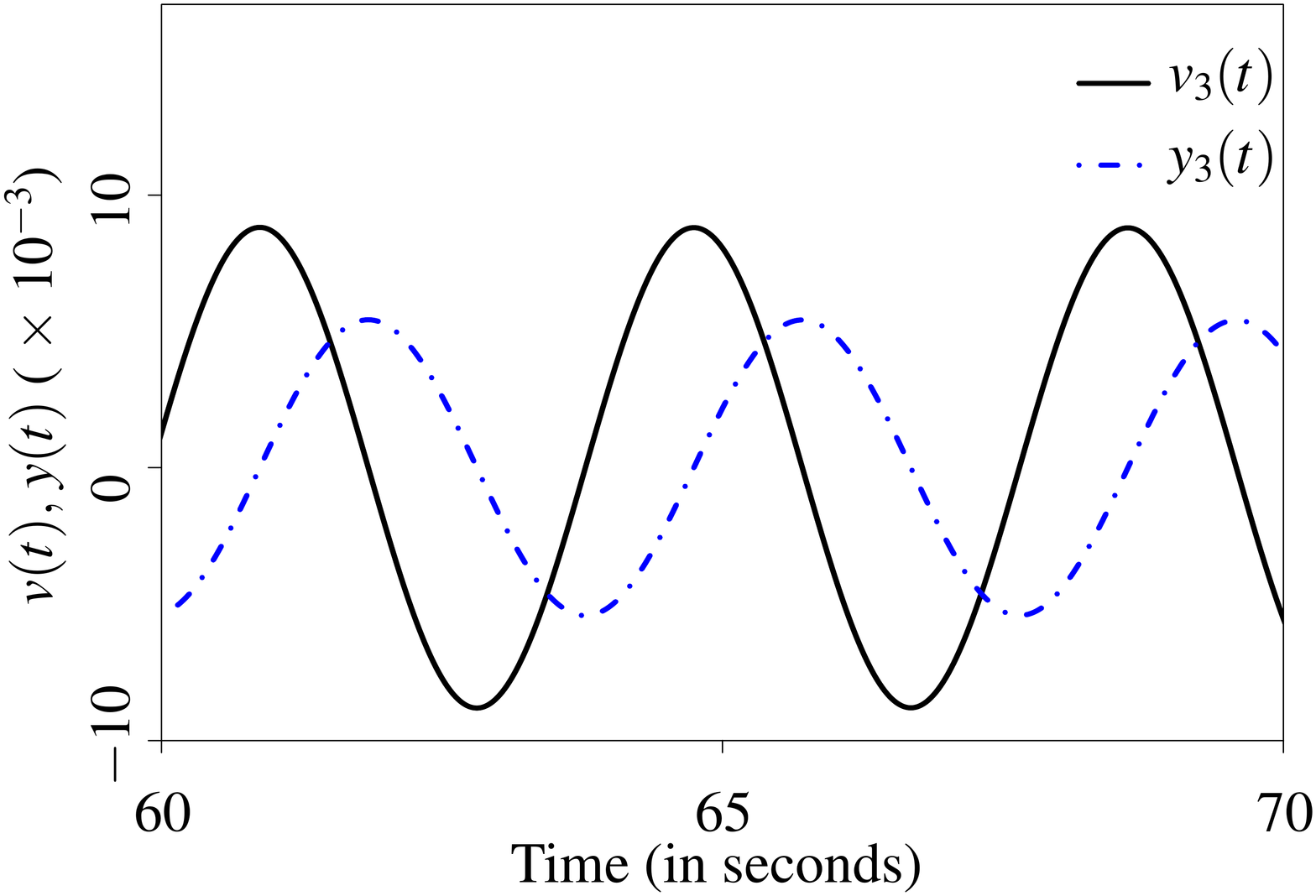}
\caption{\emph{Simulations:} Emergence of limit cycles in the headway and the relative velocity, as predicted by the analysis.}\label{fig:sims}
\end{center}
\end{figure}

We now present the simulation results for the CCFM, that serve to corroborate our analytical findings. We make use of the scientific computation software MATLAB to implement a discrete version of system~\eqref{eq:CCFMT12} with update time $T_s = 0.01$ s, thus simulating the CCFM.

We initialize the parameters with the following values. $N = 4,$ $\alpha_1 = 0.5,$ $\alpha_2 = 0.6,$ $\alpha_3 = 0.7,$ $\alpha_4 = 0.8,$ $\tau_1 = 0.5,$ $\tau_2 = 0.4,$ $\tau_3 = \tau_{cr} \approx 0.45$ and $\tau_4 = 0.3.$ The leader's velocity profile is considered to be $10 (1 - e^{-10t}),$ thus ensuring an equilibrium velocity of $10.$ Further, we fix $m = 2,$ $l = 1$ and desired headways $b_i = 20$ $\forall i.$ Fig.~\ref{fig:sims} shows the emergence of limit cycles in the state variables of the third vehicle, as predicted by our analysis. Also notice the phase shift between the relative velocity and headway solutions, as a consequence of obtaining the latter by integrating the former.

\begin{figure*}[t]
\begin{center}
\subfloat[]{
\includegraphics[scale=0.23,angle=90]{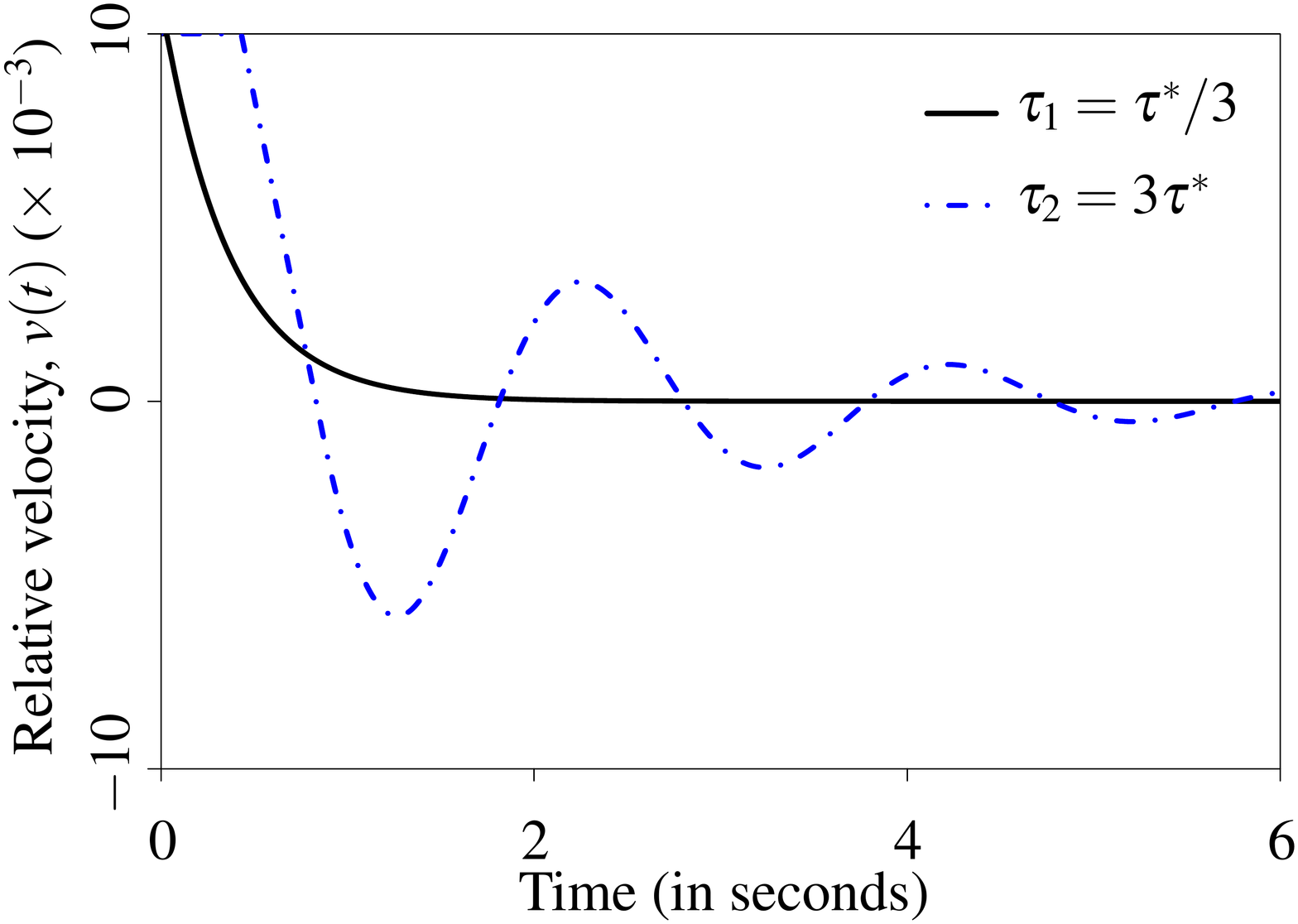}
\label{fig:nocone}
} \hspace*{20mm}
\subfloat[]{
\includegraphics[scale=0.23,angle=90]{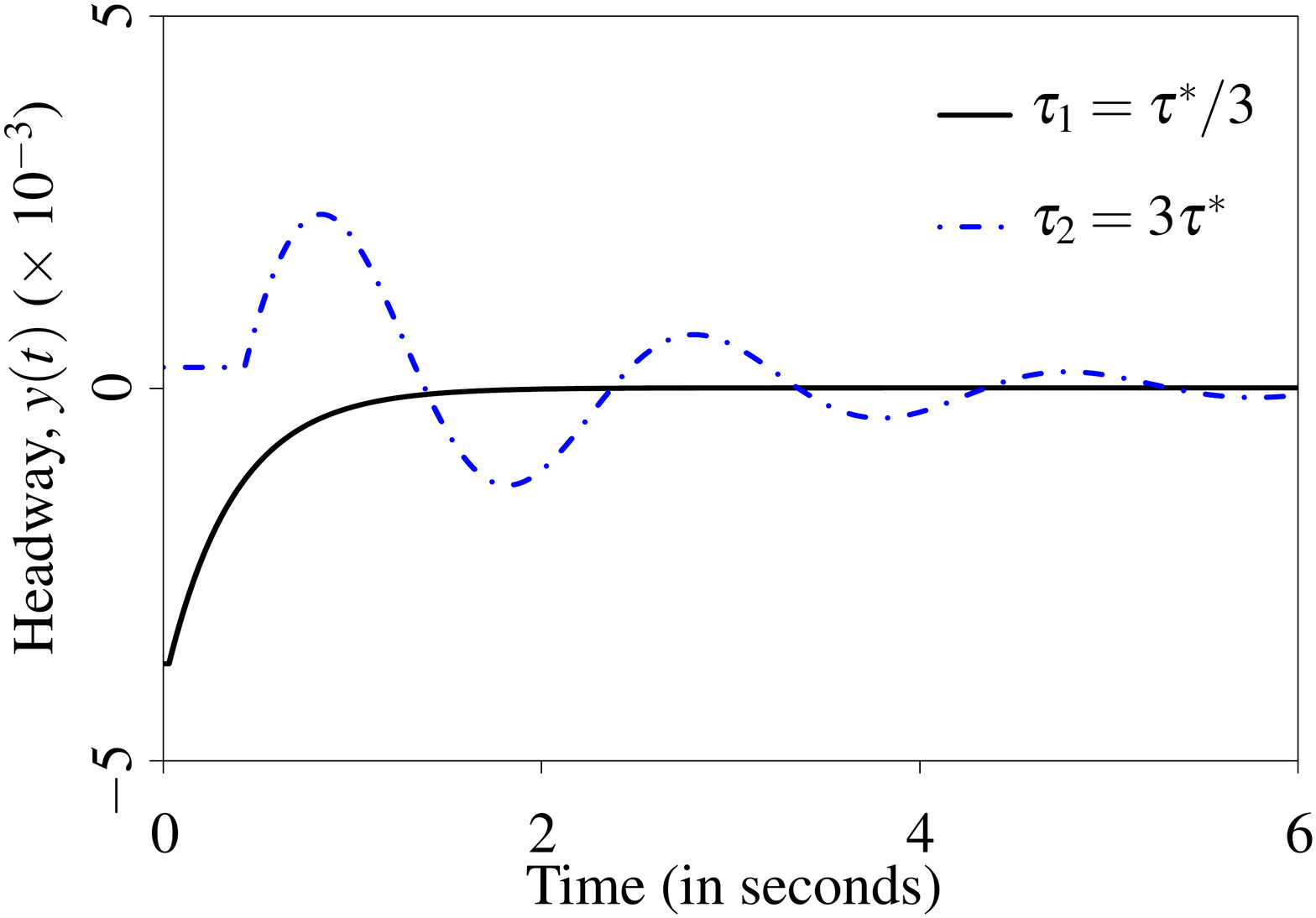}
\label{fig:noctwo}
}
\caption{\emph{Simulations:} Non-oscillatory and oscillatory solutions of the CCFM. $(a)$ portrays the relative velocity solutions, whereas $(b)$ shows the headway solutions. These serve to validate our analytical insight.}\label{fig:noc}
\end{center}
\end{figure*}

Next, we validate that the CCFM does indeed exhibit non-oscillatory convergence to the all-zero equilibrium, as predicted by~\eqref{eq:LRCCFMTnonoscil}. We also compare the rate of convergence when the reaction delay satisfies $\tau < \tau^*$ and $\tau > \tau^*,$ as discussed in Section~\ref{sec:roc}. We make use of the same parameter values as above, except for reaction delays; these are identically set to $\tau_1 = 1/(3 e \beta^*)$ and $\tau_2 = 3/( e \beta^*).$ Fig.s~\ref{fig:nocone} and~\ref{fig:noctwo} portray the solutions for relative velocity and headway respectively. To ensure comparison with Fig.~\ref{fig:rocone}, these plots correspond to the third vehicle, \emph{i.e.,} $\alpha = 0.7$  $\text{s}^{-1}$ and $l = 1.$ Notice that $\tau_1$ and $\tau_2,$ which are not in the vicinity of $\tau^*$ corresponding to the third vehicle, as seen from Fig.~\ref{fig:rocone}. Hence, the solutions corresponding to $\tau_1$ attain their equilibria much faster than those pertaining to $\tau_2.$

%% file: sec10_conc.tex
In this paper, we highlighted the importance of delayed feedback in determining the qualitative dynamical properties of a platoon of vehicles driving on a straight road. Specifically, we analyzed the Classical Car-Following Model (CCFM) in three regimes -- no delay, small delay and arbitrary delay. Control-theoretic analyses helped us derive conditions for its local stability. In particular, the analysis for small-delay regime yielded a sufficient condition for the local stability of the CCFM, whereas we obtained the necessary and sufficient condition for the local stability of the CCFM in the arbitrary-delay regime.

We then proved that the CCFM undergoes a loss of stability via a Hopf bifurcation. Mathematically, this result proves the emergence of limit cycles, which physically manifests as a back-propagating congestion wave. Even though the parameters are not strictly controllable in the case of human drivers, our work enhances the phenomenological insights into `phantom jams.' Our analyses made use of an exogenous, non-dimensional parameter that served to handle the complex relation which could exist among the various model parameters. 

We then derived the necessary and sufficient condition for non-oscillatory convergence of the CCFM. Designing control algorithms that conform to this condition ensures that jerky vehicular motions are avoided, thus guaranteeing smooth traffic flow and improving ride quality. Next, we characterized the rate of convergence of the CCFM, and highlighted the three-way trade-off between local stability, non-oscillatory convergence and the rate of convergence.

Finally, we characterized the type of Hopf bifurcation and the asymptotic orbital stability of the limit cycles, which emerge when the stability conditions are just violated, using  Poincar\'{e} normal forms and the center manifold theory. The analyses were complemented by stability charts, numerically constructed bifurcation diagrams and MATLAB simulations. These serve to highlight the impact of various model parameters on system stability as well as the relative velocity amplitude of the emergent limit cycles.

\subsection*{Avenues for further research}

There are numerous avenues that merit further investigation. In the context of the CCFM, we have addressed the issue of pairwise stability of vehicles in this work. However, string stability of a platoon of vehicles running the CCFM remains to be studied. Also, from a practical standpoint, the parameters of the CCFM may vary, for varied reasons. Hence, it becomes imperative that the longitudinal control algorithm be robust to such parameter variations, and to unmodeled vehicular dynamics.